\documentclass[pre,showpacs,amssymb,eqsecnum]{revtex4}

\usepackage{graphicx}
\begin{document}
\title{
Domain wall propagation and nucleation in a metastable two-level
system }
\author{Hans C. Fogedby}
\email{fogedby@phys.au.dk}
\affiliation
{
Institute of Physics and Astronomy,
University of Aarhus, DK-8000, Aarhus C, Denmark\\
and\\
NORDITA, Blegdamsvej 17, DK-2100, Copenhagen {\O}, Denmark
}
\author{John Hertz}
\email{hertz@nordita.dk}
\affiliation { NORDITA, Blegdamsvej 17,\\
DK-2100, Copenhagen {\O}, Denmark }
\author{Axel Svane}
\email{svane@phys.au.dk}
\affiliation
{
Institute of Physics and Astronomy,
University of Aarhus,\\
 DK-8000, Aarhus C, Denmark
}
\date{\today}
\begin{abstract}
We present a dynamical description and analysis of non-equilibrium
transitions in the noisy one-dimensional Ginzburg-Landau equation
for an extensive system based on a weak noise canonical phase
space formulation of the Freidlin-Wentzel or Martin-Siggia-Rose
methods. We derive propagating nonlinear domain wall or soliton
solutions of the resulting canonical field equations with
superimposed diffusive modes. The transition pathways are
characterized by the nucleations and subsequent propagation of
domain walls. We discuss the general switching scenario in terms
of a dilute gas of propagating domain walls and evaluate the
Arrhenius factor in terms of the associated action. We find
excellent agreement with recent numerical optimization studies.
\end{abstract}

\maketitle

\section{\label{intro}Introduction}
Phenomena far from equilibrium are ubiquitous, including
turbulence in fluids, interface and growth problems, chemical and
biological systems, and problems in material science and
nanophysics. In this context the dynamics of complex systems
driven by weak noise, corresponding to rare events, is of
particular interest. The issue of different time scales does in
fact characterize many interesting processes in nature. For
instance, in the case of chemical reactions the reaction time is
often orders of magnitude larger than the molecular vibration
periods \cite{Geissler01}. The time scale separation problem is
also encountered in the case of conformational changes of
biomolecules, nucleation events during phase transitions,
switching of the magnetization in magnetic materials
\cite{Berkov98,Koch00}, and even in the case of comets exhibiting
rapid transitions between heliocentric orbits around Jupiter
\cite{Jaffe02}.

The weak noise limit, characterizing the time scale separation, is
associated with the strong coupling regime, and the problem of
determining kinetic pathways and transition probabilities between
metastable states in systems with many degrees of freedom presents
one of the most important and challenging tasks in many areas of
physics \cite{Haenggi90}. The long time scales are associated with
the separation in energy scales of the thermal energy and the
energy barriers between metastable states; the transition takes
place by sudden jumps between metastable states followed by long
waiting times in the vicinity of the states. The fundamental issue
is thus the determination of transition pathways and the
associated transition rates.

In the weak noise limit the standard Monte Carlo method or direct
simulation of the Langevin equation becomes impractical owing to
the large separation of time scales and alternative methods have
been developed. The most notable analytical approach is the
formulation due to Freidlin and Wentzel which yields the
transition probabilities in terms of an action functional
\cite{Freidlin98}. This approach is the analogue of the
variational principle proposed by Machlup and Onsager
\cite{Machlup53,Onsager53}, see also work by Graham et al.
\cite{Graham84,Graham90}. The Freidlin-Wentzel approach is
equivalent to the Martin-Siggia-Rose method \cite{Martin73} in the
weak noise limit of the path integral formulation
\cite{deDominicis75,deDominicis76,Baussch76,Janssen76,deDominicis78}.
In order to overcome the time scale gap various numerical methods
have also been proposed. We mention here the transition path
sampling method \cite{Bolhuis02} and optimization techniques
\cite{Ren02,E02a}.

A particularly interesting non-equilibrium problem of relevance in
the nanophysics of switches is the influence of thermal noise on
two-level systems with spatial degrees of freedom, see Refs.
\cite{Berkov98,Koch00,Garcia99} . In a recent paper by E, Ren, and
Vanden-Eijden \cite{E02}, see also Ref. \cite{E02a}, this problem
has been addressed using the Ginzburg-Landau equation driven by
thermal noise. Applying the field theoretic version of the
Onsager-Machlup functional \cite{Onsager53,Machlup53} in the
Freidlin-Wentzell formulation \cite{Freidlin98}, these authors
implement a powerful numerical optimization techniques for the
determination of the space-time configuration minimizing the
Freidlin-Wentzell action and in this way determine the kinetic
pathways and their associated action, yielding the switching
probabilities in the long time-low temperature limit. The picture
that emerges from this numerical study is that of noise-induced
nucleation and subsequent propagation of domain walls across the
sample, giving rise to the switch between metastable states.

In recent work we have addressed a related problem in
non-equilibrium physics, namely the Kardar-Parisi-Zhang equation
or equivalent noisy Burgers equation describing, for example, a
growing interface in a random environment. Using a canonical phase
space method derived from the weak noise limit of the
Martin-Siggia-Rose functional or directly from the Fokker-Planck
equation, we have, in the one dimensional case, analyzed the
coupled field equations minimizing the action both analytically
\cite{Fogedby95,Fogedby98b,Fogedby98c,Fogedby01a,Fogedby01b,Fogedby03b}
and numerically \cite{Fogedby02a}. The picture that emerges is
that the transition probabilities in the weak noise limit are
associated with soliton propagation and nucleation resulting from
soliton collisions.

In the present paper we apply the canonical phase space approach
to the noisy Ginzburg-Landau equation discussed by E, Ren, and
Vanden-Eijden \cite{E02} and attempt to account for some of their
numerical findings. We thus give analytical arguments for the
propagation of noise-induced solitons and the nucleation events
originating from soliton annihilation and creation. The paper is
organized in the following way. In Sec. II we introduce the noisy
Ginzburg-Landau equation. In Sec. III we review the canonical
phase space approach. In Sec. IV we discuss diffusive mode and
domain wall solutions of the field equations replacing the noisy
Ginzburg-Landau equation. In Sec. V we analyze the domain wall
dynamics. In Sec. VI we present a stochastic interpretation of our
results. In Sec. VII we discuss equilibrium properties and kinetic
transitions. In Sec. VIII we present an interpretation of the
numerical results obtained by the optimization studies of E et. al
\cite{E02}. Sec. IX is devoted to a summary and a conclusion. A
brief version of the present work has appeared in Ref.
\cite{Fogedby03a}.
\section{\label{model}The noisy Ginzburg - Landau model}
\begin{figure}
\includegraphics[width=0.5\hsize]{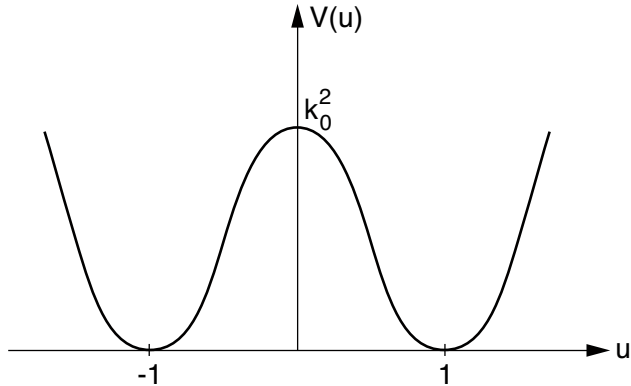}
\caption{We depict the double-well potential defining the
unperturbed states of the switch. The potential has minima at
$u=\pm 1$ and a maximum at $u=0$. } \label{fig1}
\end{figure}

The noisy Ginzburg-Landau equation driven by white noise has the
form
\begin{eqnarray}
\frac{\partial u}{\partial t} = -\Gamma\frac{\delta F}{\delta u}
+\eta, \label{gl1}
\end{eqnarray}
where the locally correlated Gaussian white noise is specified by
the moment
\begin{eqnarray}
\langle\eta(x,t)\eta(0,0)\rangle =\Delta\delta(x)\delta(t),
\label{noise}
\end{eqnarray}
with noise strength parameter $\Delta$. The free energy $F$
providing the deterministic drive is given by
\begin{eqnarray}
F=\frac{1}{2}\int dx\left(\left(\frac{\partial u}{\partial
x}\right)^2+V(u)\right). \label{free}
\end{eqnarray}
In the switching problem considered by E et al. \cite{E02} the
double well potential $V(u)$ has the form
\begin{eqnarray}
V(u) = k_0^2(1-u(x)^2)^2, \label{pot}
\end{eqnarray}
with strength parameter $k_0$. The potential vanishes at the two
minima $u\pm 1$ and assumes the maximum value $V(0)=k_0^2$ at
$u=0$. The time scale is set  by the kinetic transport coefficient
$\Gamma$ and the dimensionless scalar field $u$ determines the
spatial and temporal state of the switch. The ``Mexican hat''
double well potential is depicted in Fig.~\ref{fig1}

The Ginzburg-Landau equation in its deterministic form has been
used both in the context of phase ordering kinetics \cite{Bray94b}
and in its complex form in the study of pattern formation
\cite{Cross94}. In the noisy case for a finite system the equation
has been studied in \cite{Maier01}; see also an analysis of the
related $\phi^4$ theory in \cite{Habib00}. In the present problem
the noisy equation provides a generalization of the classical
Kramers problem \cite{Haenggi90} to spatially extended systems.

By inspection of Eqs. (\ref{gl1}), (\ref{noise}), and (\ref{free})
we note that $k_0$ has the dimension of an inverse length,
$\Gamma$ the dimension of length squared over time, and $\Delta$
the dimension of velocity. For large $k_0$, which is the case
considered here, and for $\eta=0$ the potential term dominates the
diffusive term and the field $u$ locks on to the values $\pm 1$ in
bulk; imposed boundary values being accommodated over a saturation
or healing length of order $k_0^{-1}$. In the presence of noise
the minima become unstable to thermal fluctuations and the noise
can drive the system over the potential barrier.

\begin{figure}
\includegraphics[width=0.5\hsize]{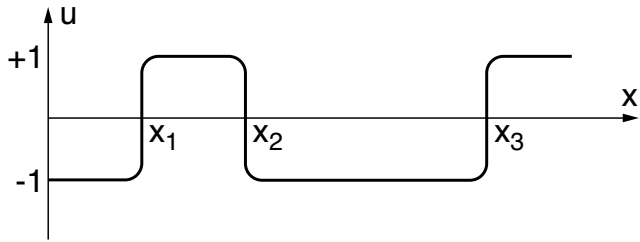}
\caption{We depict a static 3-domain wall configuration connecting
the ground states $u=-1$ and $u=+1$. The domain walls are located
at $x_1$, $x_2$, and $x_3$. The configuration corresponds to a
saddle point in the free energy landscape and has free energy
$4k_0$.} \label{fig2}
\end{figure}

The Ginzburg-Landau equation (\ref{gl1}) admits a
fluctuation-dissipation theorem yielding the stationary
distribution
\begin{eqnarray}
P_{\text{stat}}\propto\exp\left[-\frac{2\Gamma}{\Delta}F\right].
\label{stat1}
\end{eqnarray}
In a thermal environment at temperature $T$ we have
$\Delta=2\Gamma T$ (in units such that $k_{\text{B}}=1$) and
$P_{\text{stat}}\propto\exp[-F/T]$, i.e., the Boltzmann
distribution. The equilibrium states follow from the condition
\begin{eqnarray}
\frac{\delta F}{\delta u}=-\nabla^2u-2k_0^2 u(1-u^2) = 0,
\label{df}
\end{eqnarray}
yielding two degenerate uniform ground states $u=\pm 1$ with free
energy $F=0$, as well as two nonuniform domain wall solutions
\begin{eqnarray}
u_{\text{dw}}(x)=\pm\tanh k_0(x-x_0) \label{dw1},
\end{eqnarray}
centered at $x_0$ and of width $k_0^{-1}$. For large $|x|$ the
domain wall solutions overlap with the uniform ground state
solutions $u=\pm 1$. The associated domain wall free energy is
\begin{eqnarray}
F_{\text{dw}}=4k_0/3. \label{freedw}
\end{eqnarray}
The minima, maxima, and saddle point structure of the complex
energy landscape of $F$ as a functional of $\{u(x)\}$ are inferred
from the spectrum of the differential operator
\begin{eqnarray}
\frac{\delta^2F}{\delta u^2}=-\nabla^2-2k_0^2(1-3u^2). \label{d2f}
\end{eqnarray}
For the ground states $u=\pm 1$ the spectrum of $\delta^2F/\delta
u^2$ is given by the plane wave mode $\sim\exp(ikx)$ with positive
eigenvalues $k^2+4k_0^2$. For a single domain wall configuration
or $n$  connected domain walls the spectrum is composed of
zero-eigenvalue translation modes (Goldstone modes) and
superimposed phase shifted plane wave modes
$\sim\exp[ikx)\exp(i\phi)$ with positive eigenvalues $k^2+4k_0^2$,
see e.g. Ref. \cite{Fogedby85}. We thus infer that the free energy
landscape possesses two global minima at $u=\pm 1$ and a series of
local saddle points of free energy $4nk_0/3$, corresponding to $n$
connected domain walls at positions $x_i, i=1,2,\cdots n$. In
Fig.~\ref{fig2} we have depicted a static 3-domain wall
configuration connecting the ground states $u=-1$ and $u=+1$ with
free energy $4k_0$.
The static single domain wall or multi-domain wall configurations
correspond to points in the free energy landscape $F=F(\{u(x)\})$
and cannot effectuate a switch between the ground state
configurations $u=\pm 1$. However, in the presence of noise the
Kramers escape mechanism sets in and the system can perform a
kinetic transition. The transition probability is typically
characterized by an Arrhenius factor $P\sim\exp[-\Gamma\delta
F/\Delta]$ where $\delta F$ is the free energy associated with the
potential barriers encountered in the dynamic transition, see e.g.
\cite{Haenggi90}. In order to address this issue for the noisy
Ginzburg-Landau equation we apply the canonical phase space
approach.
\section{\label{method}The canonical phase space approach}
The canonical phase space approach is discussed in detail in
\cite{Fogedby99a,Fogedby02a}. Briefly, the Fokker-Planck equation
for the probability distribution $P(\{u(x)\},t)$ has the general
form
\begin{eqnarray}
\Delta\frac{\partial P}{\partial t}=HP,
\label{fp}
\end{eqnarray}
with formal solution $P\propto\exp[Ht/\Delta]$. The Fokker-Planck
equation (\ref{fp}) is driven by the Hamiltonian or Liouvillian
$H(u,\delta/\delta u)$ (a differential-functional operator in
$u(x)$ space). The method is based on a weak noise WKB-like
approximation
\begin{eqnarray}
P\propto\exp\left[-\frac{S}{\Delta}\right], \label{dis1}
\end{eqnarray}
applied to Eq. (\ref{fp}). To leading order in the noise strength
$\Delta$ the action $S$ satisfies the Hamilton-Jacobi equation
\begin{eqnarray}
\frac{\partial S}{\partial t}+H(p,u)=0, \label{hj}
\end{eqnarray}
with canonical momentum $p=\delta S/\delta u$, defining the
Hamiltonian $H$ as a functional of $u$ and $p$ and implying a
principle of least action. The variational principle $\delta S
=0$, moreover, implies the Hamiltonian field equations of motion
\begin{eqnarray}
&&\frac{\partial u}{\partial t}=\frac{\delta H}{\delta p},
\label{equ1}
\\
&&\frac{\partial p}{\partial t}=-\frac{\delta H}{\delta u},
\label{eqp1}
\end{eqnarray}
and the action
\begin{eqnarray}
S=\int dxdt~p\frac{\partial u}{\partial t} - \int dt~H.
\label{act1}
\end{eqnarray}
Applying this scheme to the Ginzburg-Landau equation (\ref{gl1})
we obtain
\begin{eqnarray}
&&\frac{\partial u}{\partial t}=-\Gamma\frac{\delta F}{\delta u}
+p, \label{feu1}
\\
&&\frac{\partial p}{\partial t}= \Gamma \frac{\delta^2 F} {\delta
u^2}p, \label{fep1}
\end{eqnarray}
driven by the Hamiltonian (the generator of time translations)
\begin{eqnarray}
H=\int dx~{\cal H}=\frac{1}{2}\int dx~p\left[p-2\Gamma\frac{\delta
F}{\delta u}\right], \label{ham1}
\end{eqnarray}
with $\delta F/\delta u$ and $\delta^2F/\delta u^2$ given by Eqs.
(\ref{df}) and (\ref{d2f}). As a result the action $S$ associated
with an orbit from $u_1$ to $u_2$ traversed in time $T$ is given
by
\begin{eqnarray}
S(u_1, u_2,T)=\int_{u_1,0}^{u_2,T}dxdt~\left[p\frac{\partial
u}{\partial t} - {\cal H}\right],
\label{act2}
\end{eqnarray}
where $\cal H$ is the Hamiltonian density. In our further analysis
we shall also make use of the total momentum $\Pi$ (the generator
of space translations)
\begin{eqnarray}
\Pi=\int dx~u\frac{\partial p}{\partial x}. \label{mom1}
\end{eqnarray}
The prescription for determining the relevant Arrhenius factor
associated with a transition and the kinetic pathway is thus
straightforward. Fixing boundary values in space and time the
first step is to solve the coupled field equations (\ref{feu1})
and (\ref{fep1}) for an orbit (a minimizer in the terminology of
Ref. \cite{E02}) from $u_1$ to $u_2$ traversed in time $T$, next
we evaluate the action $S$ along an orbit according to Eq.
(\ref{act2}), and finally deduce the transition probability from
Eq. (\ref{dis1}).

The correspondence with the equivalent Freidlin-Wentzel action
$S_{\text{FW}}$ is obtained by inserting the equation of motion
(\ref{feu1}) and the Hamiltonian (\ref{ham1}) in the action
(\ref{act2}), yielding $S=(1/2)\int dxdtp^2$ or
$S_{\text{FW}}=(1/2)\int dxdt(\partial u/\partial t+\Gamma\delta
F/\delta u)^2$. The Freidlin-Wentzel method is a Lagrangian
configuration space method with Lagrangian density ${\cal
L}=(1/2)(\partial u/\partial t+\Gamma\delta F/\delta u)^2$,
whereas the present approach is a phase space formulation with
action (\ref{act2}). We also note that the equations of motion
(\ref{feu1}) and (\ref{fep1}) are identical to the saddle point
equations in the Martin-Siggia-Rose functional formulation
\cite{Martin73,Baussch76}.
\section{\label{solutions}Diffusive mode and domain wall solutions}
More explicitly, inserting $\delta F/\delta u$ and
$\delta^2F/\delta u^2$ from Eqs. (\ref{df}) and (\ref{d2f}) the
equations of motion (\ref{feu1}) and (\ref{fep1}) assume the form
\begin{eqnarray}
&&\frac{\partial u}{\partial t} = \Gamma\frac{\partial^2 u}
{\partial x^2}+2\Gamma k_0^2u(1-u^2)+p, \label{feu2}
\\
&&\frac{\partial p}{\partial t}=-\Gamma\frac{\partial^2p}
{\partial x^2}-2\Gamma k_0^2 p(1-3u^2), \label{fep2}
\end{eqnarray}
and the Hamiltonian is
\begin{eqnarray}
H=\frac{1}{2}\int dx p\left(p+2\Gamma\frac{\partial^2u}{\partial
x^2}+ 4\Gamma k_0^2 u(1-u^2)\right). \label{ham2}
\end{eqnarray}
The equations of motion (\ref{feu2}) and (\ref{fep2}) determine
orbits in a multi-dimensional phase space $\{u(x),p(x)\}$ lying on
the energy manifolds determined by Eq. (\ref{ham2}); for open or
periodic boundary conditions the orbits are moreover confined by
the conservation of the total momentum $\Pi$ given by Eq.
(\ref{mom1}). From an analytical point of view the field equations
(\ref{feu2}) and (\ref{fep2}) are in general intractable. Even
numerically, the negative diffusion term in Eq. (\ref{fep2})
renders the coupled equations highly unstable as was noted in the
numerical analysis of the Burgers equation \cite{Fogedby02a}.
\subsection{Linear diffusive modes}
\begin{figure}
\includegraphics[width=0.5\hsize]{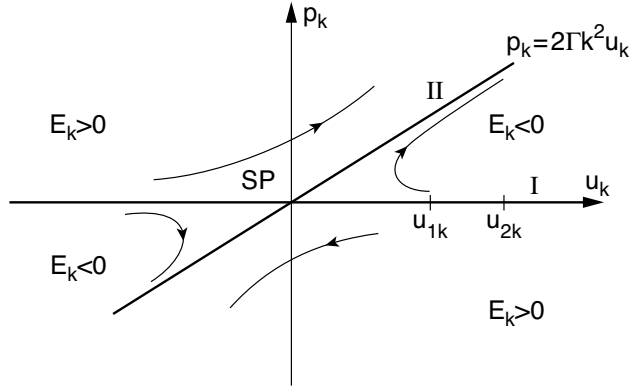}
\caption{ We show the orbits in the $(u_k,p_k)$ phase space. The
finite time orbit from $u_{1k}$ to $u_{2k}$ lies on the energy
manifold $E_k=p_k^\ast[p_k-2\Gamma k^2u_k]$. In the long time
limit $T\rightarrow\infty$ the orbits migrate to the transient
manifold $p_k=0$ (I) and the stationary manifold $p_k=2\Gamma
k^2u_k$ (II) passing through the saddle point $(u_k,p_k)=(0,0)$
(SP) implying ergodicity and a stationary state. } \label{fig3}
\end{figure}

It is instructive first to consider the easily discussed linear
case of simple diffusion for small $k_0$. Thus ignoring the
non-linear potential terms in Eqs. (\ref{feu2}) and (\ref{fep2})
we obtain the linear equations
\begin{eqnarray}
&&\frac{\partial u}{\partial t} =
\Gamma\frac{\partial^2u}{\partial x^2}+p, \label{linu1}
\\
&&\frac{\partial p}{\partial t}=
-\Gamma\frac{\partial^2p}{\partial x^2}, \label{linp1}
\end{eqnarray}
generated by the Hamiltonian
\begin{eqnarray}
H=\frac{1}{2}\int dx p\left[p+2\Gamma\frac{\partial^2u}{\partial
x^2}\right]. \label{ham3}
\end{eqnarray}
In Fourier space, setting $u_k=\int dx~\exp{(-ikx)}u$ and
$p_k=\int dx~\exp(-ikx)p$, $u_k^\ast=u_{-k}$ and $p_k^\ast =
p_{-k}$, the equations of motion decompose and we arrive at the
orbit solution, see Ref. \cite{Fogedby99a},
\begin{eqnarray}
u_k(t)= \frac {u_{2k}\sinh\Gamma k^2t+u_{1k}\sinh\Gamma k^2(T-t)}
{\sinh\Gamma k^2T}, \label{solu1}
\end{eqnarray}
from $u_{1k}$ at time $t=0$ to $u_{2k}$ at time $t=T$;
$u_{1,2}=\int (dk/2\pi)\exp(ikx)u_{1,2,k}$. The noise field $p_k$
is slaved to the motion of $u_k$ and given by
\begin{eqnarray}
p_k(t)= \Gamma k^2 e^{\Gamma k^2t}\frac{u_{2k}-u_{1k}e^{-\Gamma
k^2 T}}{\sinh\Gamma k^2T}. \label{solp1}
\end{eqnarray}
Likewise, the Hamiltonian $H$ decomposes into independent $k$-mode
contributions
\begin{eqnarray}
H=\int\frac{dk}{2\pi}p_k^\ast[p_k-2\Gamma k^2u_k]. \label{ham4}
\end{eqnarray}
Inserting the solutions (\ref {solu1}) and (\ref{solp1}) we obtain
specifically for the energy of the $k$-th mode
\begin{eqnarray}
E_k=\frac{(\Gamma k^2)^2}{2}\frac{
|u_{2k}|^2+|u_{1k}|^2-2u_{1k}u_{2k}\cosh\Gamma k^2T}
{\sinh^2\Gamma k^2 T }. \label{e1}
\end{eqnarray}
The orbits lie on the energy manifolds given by
$E_k=p_k^\ast[p_k-2\Gamma k^2u_k]$. In the long time limit
$T\rightarrow\infty$ the orbits migrate to the zero-energy
manifolds consisting of the transient submanifold $p_k=0$ and the
stationary submanifold $p_k=2\Gamma k^2 u_k$, and pass
asymptotically through the hyperbolic fixed point at
$(u_k,p_k)=(0,0)$, determining the stationary state. In
Fig.~\ref{fig3} we have shown the orbits for a particular $k$-mode
in a plot of $p_k$ versus $u_k$.

Finally, the conserved momentum $\Pi$ and the action $S$ follow
from Eqs. (\ref{mom1}) and (\ref{act2})
\begin{eqnarray}
\Pi = \int\frac{dk}{2\pi}(-ik)p_k^\ast u_k, \label{mom2}
\end{eqnarray}
and
\begin{eqnarray}
S = \int \frac{dk}{2\pi}\Gamma k^2\frac{|u_{2k}-u_{1k}e^{-\Gamma
k^2T}|^2} {1-e^{-2\Gamma k^2 T}}, \label{act3}
\end{eqnarray}
yielding the transition probability from $u_{1k}$
to $u_{2k}$ in time $T$,
\begin{eqnarray}
P(u_{1k},u_{2k},T)\propto \exp\left[ -\frac{1}{\Delta} \int
\frac{dk}{2\pi} \Gamma k^2 \frac{|u_{2k}-u_{1k}e^{-\Gamma k^2
T}|^2} {1-e^{-2\Gamma k^2 T}} \right]. \label{dis2}
\end{eqnarray}
In the long time limit $T\rightarrow\infty$ we reach the
stationary distribution
\begin{eqnarray}
P_{\text{st}}(\{u_k\})\propto\exp\left[-\frac{1}{\Delta} \int
\frac{dk}{2\pi}\Gamma k^2|u_k|^2\right]. \label{stat2}
\end{eqnarray}
This result is consistent with the noise driven diffusion equation
written in the form $\partial u/\partial t = -\Gamma\delta
F/\delta u+\eta$,
$\langle\eta\eta\rangle=\Delta\delta(x)\delta(t)$, with free
energy $F = (1/2)\int dx(\partial u/\partial x)^2$. Invoking the
fluctuation-dissipation theorem we then obtain an equilibrium
distribution given by the Boltzmann factor $\exp(-F/T)$,
$\Delta=2\Gamma T$, consistent with Eq. (\ref{stat2}).

In summary, in the linear purely diffusive case the configurations
$u(x,t)$ decompose into independent $k$-modes. In the dynamical
phase space approach the noise $\eta_k$ driving the individual
$k$-modes is replaced by the noise field $p_k$ which couples
parametrically to the time evolution of the $u_k$ mode. In a
transient time regime $u_k$ is damped according to the diffusion
equation and the orbit in phase space lies close to the transient
or noiseless submanifold $p_k=0$ and approaches the saddle point
$(u_k,p_k)=(0,0)$. At longer times the growing noise field $p_k$
drives $u_k$ away from the saddle point and the orbit approaches
the stationary or noisy submanifold $p_k=2\Gamma k^2 u_k$, i.e.,
the distribution associated with the noise-driven diffusion
equation approaches a stationary distribution. In the limit
$T\rightarrow\infty$ the orbit passes asymptotically through the
saddle point and the orbit from $u_{1k}$ to $u_{2k}$ lies on the
zero-energy manifolds determining the stationary state.

\subsection{Domain wall modes}
In the nonlinear case the phase space representation of the
noise-driven Ginzburg Landau equation is given by the coupled
field equations (\ref{feu2}) and (\ref{fep2}) determining the
orbits. As in the linear case we can identify the zero-energy
submanifolds determining the stationary state. This is related to
the existence of a fluctuation-dissipation theorem for the noisy
Ginzburg-Landau equation (\ref{gl1}) expressed by the existence of
a free energy. From the Hamiltonian (\ref{ham1}) we infer
tentatively the zero-energy submanifolds $p=0$ and
$p=2\Gamma\delta F/\delta u$. The transient noiseless submanifold
$p=0$ is consistent with the equations of motion and the
configurations decay according to the damped deterministic
Ginzburg-Landau equation
\begin{eqnarray}
\frac{\partial u}{\partial t}=-\Gamma\frac{\delta F}{\delta u}.
\label{gl2}
\end{eqnarray}
Note that in contrast to the linear case where the $u$
configurations decompose in $k$-modes which decay according to
$u_k\propto\exp(-\Gamma k^2t)$, an initial configuration $u_1$ in
the nonlinear case will in general decay forming a pattern of
interacting and annihilating domain wall configurations with
superimposed diffusive modes.

The stationary submanifold $p=2\Gamma\delta F/\delta u$ inserted
in Eqs. (\ref{feu1}) and (\ref{fep1}) yields the equations
\begin{eqnarray}
&&\frac{\partial u}{\partial t} = \Gamma\frac{\delta F}{\delta u},
\label{equ2}
\\
&&\frac{\partial p}{\partial t}=\Gamma^2\frac{\delta}{\delta
u}\left(\frac{\delta F}{\delta u}\right)^2, \label{eqp2}
\end{eqnarray}
which are consistent since $\partial u/\partial t=(1/2)p$. It is
interesting to notice that the motion of $u$ on the stationary
noisy submanifold $p=2\Gamma(\delta F/\delta u)$ is a
time-reversed version of the motion on the transient submanifold
$p=0$. Finally, on the zero-energy manifold the action in Eq.
(\ref{act1}) takes the form $S=\int dxdt~p\partial u/\partial
t=2\Gamma\int dxdt~(\partial u/\partial t)(\delta F/\delta u) =
2\Gamma F$, yielding the stationary distribution in Eq.
(\ref{stat1}) for the noisy Ginzburg-Landau equation.

In the absence of a fluctuation-dissipation theorem as is for
example the case for the kinetic KPZ equation or, equivalently,
noisy Burgers equation, we can in general not explicitly identify
the stationary zero-energy submanifold and thus simply determine
the stationary state; an exception is the KPZ equation in 1D where
for special reasons a fluctuation-dissipation theorem is
available, see Refs. \cite{Fogedby98b,Fogedby98c,Fogedby99a}.

In  order to address non-equilibrium properties such as a specific
transition probability from an initial state $u_1$ to a final
state $u_2$ in passage time $T$ we must address the nonlinear
equations of motion (\ref{feu2}) and (\ref{fep2}). The field
equations are not integrable and do not yield to a general
analytical solution. However, we can advance our understanding by
first searching for static solutions on the transient manifold
$p=0$, i.e., the solution of the equation $\delta F/\delta u=0$
yielding according to Eq. (\ref{df}) the domain wall solutions in
Eq. (\ref{dw1}). The domain wall excitations are of the instanton
type and can be located at arbitrary positions, see Ref.
\cite{Rajaraman87}. Since the overlap between two well-separated
domain walls is exponentially small we can construct approximate
multi-domain wall solutions of the form
\begin{eqnarray}
&&u_{\text{dw}}=\sum_{i=1}^n\sigma_i\tanh k_0(x-x_i)+u_0,
\label{dwu2}
\\
&&p_{\text{dw}}=0. \label{dwp2}
\end{eqnarray}
Here the parity index $\sigma_i=\pm 1$ for right hand and left
hand domain walls, respectively, and $x_i$ indicates the center of
the domain wall. The offset $u_0=0$ for multi-domain wall
configurations overlapping for large $|x|$ with two different
ground state configurations and $u_0=\pm 1$ for configurations
overlapping with identical ground states $u=\pm 1$, respectively.
Assuming that the inter-domain wall distance $|x_i-x_{i+1}|$ is
large compared with the domain wall width $1/k_0$, i.e., the case
of a dilute domain wall gas, the expression (\ref{dwu2})
constitutes an approximate solution to Eq. (\ref{df}). Since the
domain wall solutions are associated with the transient
submanifold $p=0$ it also follows from Eqs. (\ref{ham2}),
(\ref{mom1}) and (\ref{act2}) that they carry vanishing energy,
momentum, and action within the canonical phase space approach.

\section{\label{dynamics}Domain wall dynamics}
In order to impart dynamical attributes to the domain walls and
thus provide solutions to the coupled field equations (\ref{feu2})
and (\ref{fep2}) we perform a linear stability analysis about the
static domain wall solutions.
\subsection{Dynamics of a single domain wall}
Setting $u=u_{\text{dw}}+\delta u$ and $p=p_{\text{dw}}+\delta p$,
$p_{\text{dw}}=0$ in Eqs. (\ref{feu2}) and (\ref{fep2}) we obtain
to linear order the coupled equations
\begin{eqnarray}
&&\frac{\partial\delta u}{\partial t}=\Gamma\frac{\partial^2\delta
u}{\partial x^2}+2\Gamma k_0^2(1-3u^2_{\text{dw}})\delta u+\delta
p, \label{linu2}
\\
&&\frac{\partial\delta p}{\partial
t}=-\Gamma\frac{\partial^2\delta p}{\partial x^2}-2\Gamma
k_0^2(1-3u^2_{\text{dw}})\delta p. \label{linp2}
\end{eqnarray}
Noting that in a Schr\"{o}dinger equation analogue the domain wall
profile $u_{\text{dw}}$ gives rise to a Bargmann type potential
Eqs. (\ref{linu2}) and (\ref{linp2}) are readily analyzed, see
e.g. Ref. \cite{Fogedby85}. Expanding $\delta u$ and $\delta p$ on
the eigenfunctions $\Psi_n$ of the Schr\"{o}dinger operator
\begin{eqnarray}
D=-\nabla^2+2k_0^2\left[2-\frac{3}{\cosh^2k_0x}\right],
\label{op1}
\end{eqnarray}
according to
\begin{eqnarray}
&&\delta u=\sum_nu_n\Psi_n, \label{delu1}
\\
&&\delta p=\sum_np_n\Psi_n, \label{delp1}
\end{eqnarray}
the time-dependent expansion coefficients $u_n$ and $p_n$ are
determined by the coupled equations of motion
\begin{eqnarray}
&&\frac{d u_n}{dt}= -\Gamma\lambda_n u_n + p_n, \label{linu3}
\\
&&\frac{d p_n}{dt}=\Gamma\lambda_np_n, \label{linp3}
\end{eqnarray}
where $\lambda_n$ is the eigenvalue in the eigenvalue equation
$D\Psi_n=\lambda_n\Psi_n$. Equations (\ref{linu3}) and
(\ref{linp3}) have the same structure as the Fourier transformed
versions of Eqs. (\ref{linu1}) and (\ref{linp1}) in the case of
the linear diffusive modes, and the solutions are given by Eqs.
(\ref{solu1}) and (\ref{solp1}) with $\Gamma k^2$ replaced by
$\Gamma\lambda_n$. The canonical phase space structure with
$\Gamma k^2$ replaced by $\Gamma\lambda_n$ is depicted in
Fig.~\ref{fig3}.

The spectrum of $D$ in Eq. (\ref{op1}) is composed of two bound
states $\Psi_0$ and $\Psi_1$ with eigenvalues $\lambda_0=0$ and
$\lambda_1=3k_0^2$, respectively, and a band of phase-shifted
plane wave solutions $\Psi_k$ with eigenvalues
$\lambda_k=k^2+4k_0^2$
\begin{eqnarray}
&&\Psi_0=\left(\frac{3k_0}{4}\right)^{1/2}\frac{1}
{\cosh^2k_0x},~~\lambda_0=0, \label{bs0}
\\
&&\Psi_1=\left(\frac{3k_0}{2}\right)^{1/2}\frac{\sinh
k_0x}{\cosh^2k_0x},~~\lambda_1=3k_0^2, \label{bs1}
\\
&&\Psi_k=\left(\frac{1}{2\pi}\right)^{1/2}e^{ikx}s_k(x),~~~~
\lambda_k=k^2+4k_0^2, \label{band}
\\
&&s_k(x)=\frac{k^2+k_0^2-3k_0^2\tanh^2k_0x+3ikk_0\tanh k_0x}
{(k-ik_0)(k-2ik_0)}.
\label{smatrix}
\end{eqnarray}
The complex space-dependent s-matrix $s_k(x)$ gives rise to a
space and phase modulation of the plane wave mode. For
$x\rightarrow -\infty$ we have $s_k(x)\rightarrow 1$, whereas for
$x\rightarrow\infty$ we obtain
$s_k(x)\rightarrow\exp(i\delta_0(k))\exp(i\delta_1(k))$ where the
phase shifts $\delta_0=2\tan^{-1}(k_0/k)$ and
$\delta_1=2\tan^{-1}(2k_0/k)$ are associated with the depletion of
the band due to the formation of the bound states $\Psi_0$ and
$\Psi_1$, respectively. In terms of the expansion coefficients
$u_n$ and $p_n$ the Hamiltonian $H$,  momentum $\Pi$, and  action
$S$ in Eqs. (\ref{ham2}), (\ref{mom1}), and (\ref{act2})  are
given by
\begin{eqnarray}
&&H=\frac{1}{2}\sum_np_n^\ast(p_n-2\Gamma\lambda_nu_n),
\label{ham5}
\\
&&\Pi=\sum_np_n\int dxu_{\text{dw}}\nabla\Psi_n \label{mom3}
\\
&&S=\frac{1}{2}\sum_n\int dtp_n^\ast p_n, \label{act4}
\end{eqnarray}
expressing the dynamics of the various modes.
\subsubsection{Domain wall motion}
The mode $\Psi_0$ with eigenvalue $\lambda_0=0$ plays a special
role since it is associated with the uniform translation of the
domain wall. For $\lambda_0=0$ the equation of motions
(\ref{linu3}) and (\ref{linp3}) take the form $du_0/dt=p_0$ and
$dp_0/dt=0$ with solutions $u_0=p_0t$, $p_0=\text{cst}$. For the
field $u$ associated with the mode $\Psi_0$ we have
$u=u_{\text{dw}}+u_0\Psi_0$. Inserting $\Psi_0$ and $u_0=p_0t$ we
obtain
\begin{eqnarray}
u=u_{\text{dw}}+p_0\sigma
t\left(\frac{3k_0}{4}\right)^{1/2}k_0^{-1}\nabla
u_{\text{dw}}\propto u_{\text{dw}}(x-vt), \label{dwu3}
\end{eqnarray}
describing a domain wall propagating with velocity
\begin{eqnarray}
v=-p_0\sigma\left(\frac{3}{4k_0}\right)^{1/2},~~\sigma=\pm 1.
\label{vel1}
\end{eqnarray}
The mode $\Psi_0$ is the well-known translation or Goldstone mode
restoring the broken translation symmetry associated with the
localized domain wall mode. In the present canonical formulation
the translation mode also implies the propagation of the domain
wall.

\begin{figure}
\includegraphics[width=.5\hsize]{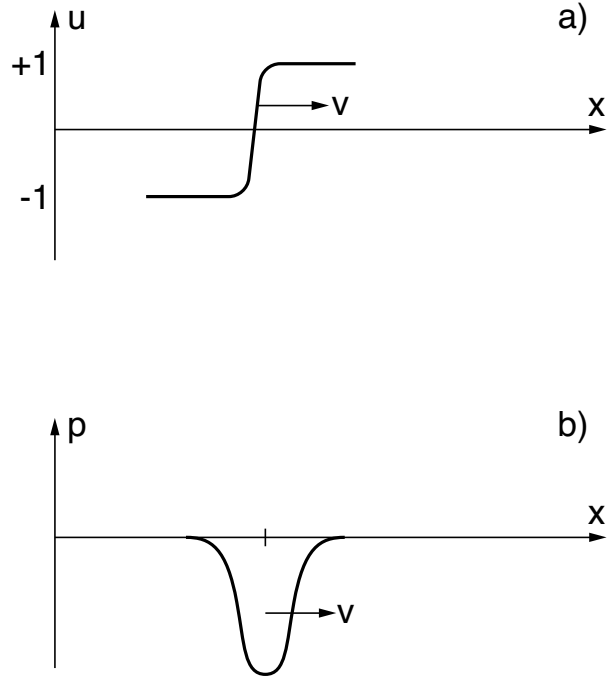}
\caption{In a) we show a right hand domain wall  in $u$ moving
with velocity $v$. In b) we depict the associated impulsive noise
field $p$ } \label{fig4}
\end{figure}

It is also instructive to consider the noise field $p$ which in
the dynamical description corresponds to the noise $\eta$. In the
case of a single domain wall we obtain from Eq. (\ref{delp1}) a
static configuration. It is, however, clear that the noise field
must move together with the domain wall configurations and we
conclude that terms beyond linear order give rise to a
renormalization of the noise field profile implying a finite
propagation velocity, i.e.,
\begin{eqnarray}
p\propto p_0\frac{1}{\cosh^2k_0(x-vt)}. \label{solp}
\end{eqnarray}
The noise field is thus localized at the position of the domain
wall corresponding to a noise impulse associated with the
formation of the domain wall. In Fig.~\ref{fig4} we have depicted
a single moving right hand domain wall (index $\sigma=1$) with
associated noise field.
The dynamics of a single domain wall follows from Eqs.
(\ref{ham5}), (\ref{mom3}), and (\ref{act4}); we have
\begin{eqnarray}
&&E_0=\frac{1}{2}p_0^2, \label{e2}
\\
&&\Pi_0=-p_0\sigma\left(\frac{4k_0}{3}\right)^{1/2}, \label{m2}
\\
&&S_0=\frac{1}{2}Tp_0^2. \label{s2}
\end{eqnarray}
In terms of the propagation velocity $v$ we have $\Pi_0=(4/3)k_0v$
and we can associate an effective mass $(4/3)k_0$ with the domain
wall motion,
\begin{eqnarray}
m=\frac{4}{3}k_0;
\end{eqnarray}
note that the mass vanishes in the limit of a broad domain wall.
\subsubsection{Deformation and extended modes}
The bound state $\Psi_1$ given by (\ref{bs1}) is odd and accounts
for the symmetrical deformation of the moving domain wall. The
time-dependence of $u$ is given by Eqs. (\ref{solu1}) and
(\ref{solp1}) with $\Gamma k^2$ replaced by
$\Gamma\lambda_1=3\Gamma k_0^2$. The band of plane wave solutions
in Eq. (\ref{band}) corresponds to corrections to the leading and
trailing edges of the domain wall. The modes have diffusive
character with growing and damped time behavior and are given by
Eqs. (\ref{solu1}) and (\ref{solp1}) with $\Gamma k^2$ replaced by
$\Gamma(k^2+4k_0^2)$; note that in the limit $k_0\rightarrow 0$ we
recover the linear case. The extended modes are phase-shifted
$2\tan^{-1}(k_0/k)+2\tan^{-1}(2k_0/k)$ which by Levinson's theorem
corresponds to the two bound states, the translation mode and
deformation mode, being depleted from the band.
\subsection{Dynamics of multi-domain wall configurations}
In the multi-domain wall case the analysis proceeds in a similar
manner. Expanding about the static multi-domain wall configuration
in Eqs. (\ref{dwu2}) and (\ref{dwp2}), $u=u_{\text{dw}}+\delta u$
and $p=p_{\text{dw}}+\delta p$, $p_{\text{dw}}=0$, we obtain for a
dilute domain wall gas the coupled linear equations
\begin{eqnarray}
&&\frac{\partial\delta u}{\partial t}=\Gamma\frac{\partial^2\delta
u}{\partial x^2}+ 2\Gamma k_0^2
\left[-2+3\sum_{i=1}^n\frac{1}{\cosh^2k_0(x-x_i)}\right]\delta u
+\delta p, \label{linu4}
\\
&&\frac{\partial\delta p}{\partial t} =
-\Gamma\frac{\partial^2\delta p}{\partial x^2}- 2\Gamma k_0^2
\left[-2+3\sum_{i=1}^n\frac{1}{\cosh^2k_0(x-x_i)}\right]\delta p,
\label{linp4}
\end{eqnarray}
which are analyzed in terms of the spectrum of the Schr\"{o}dinger
operator
\begin{eqnarray}
D=-\nabla^2+2k_0^2
\left[2-3\sum_{i=1}^n\frac{1}{\cosh^2k_0(x-x_i)}\right],
\label{op2}
\end{eqnarray}
with identical well-separated potential wells at $x_i$. Expanding
$\delta u$ and $\delta p$ on the eigenfunctions $\Psi_n$ of $D$,
$D\Psi_n=\lambda_n\Psi_n$, we recover the equations of motion
(\ref{linu3}) and (\ref{linp3}).

The eigenstates $\Psi_n$ are readily expressed as linear
superpositions of the eigenstates for the individual potential
wells and we obtain the translation modes $\Psi_0$ with eigenvalue
$\lambda_0=0$, the deformation modes $\Psi_1$ with eigenvalue
$\lambda_1=3k_0^2$, and the extended plane wave modes $\Psi_k$
with eigenvalues $\lambda_k=k^2+4k_0^2$,
\begin{eqnarray}
&&\Psi_0=\sum_{i=1}^n A^i_0\Psi_0^i,~~~~~~~~~~~~~~
\Psi_0^i\propto\frac{1}{\cosh^2k_0(x-x_i)}, \label{mbs0}
\\
&&\Psi_1=\sum_{i=1}^n A^i_1\Psi_1^i,~~~~~~~~~~~~~~
\Psi_1^i\propto\frac{\sinh k_0(x-x_i)}{\cosh^2k_0(x-x_i)}~,
\label{mbs1}
\\
&&\Psi_k\propto e^{ikx} \prod_{i=1}^n s_k(x-x_i),~~
s_k(x)=\frac{k^2+k_0^2-3k_0^2\tanh^2k_0x+3ikk_0\tanh k_0x}
{(k-ik_0)(k-2ik_0)}. \label{mband}
\end{eqnarray}
In terms of the individual eigenfunctions the expansions of
$\delta u$ and $\delta p$ take the form
\begin{eqnarray}
&&\delta u=\sum_{ni}u_{ni}\Psi_n^i,~~u_{ni}=u_nA_i, \label{delu2}
\\
&&\delta p=\sum_{ni}p_{ni}\Psi_n^i,~~u_{ni}=p_nA_i, \label{delp2}
\end{eqnarray}
and we obtain the equations of motion
\begin{eqnarray}
&&\frac{d u_{ni}}{dt}=-\Gamma\lambda_nu_{ni} + p_{ni},
\label{linu5}
\\
&&\frac{d p_{ni}}{dt}=\Gamma\lambda_np_{ni}, \label{linp5}
\end{eqnarray}
with solutions given by Eqs. (\ref{solu1}) and (\ref{solp1}).
Moreover, the dynamics of a multi-domain wall configuration is
given by
\begin{eqnarray}
&&H=\frac{1}{2}\sum_{ni}p_{ni}^\ast(p_{ni}-2\Gamma\lambda_nu_{ni}),
\label{ham6}
\\
&&\Pi=\sum_{ni}p_{ni}\int dxu_{\text{dw}}\nabla\Psi_n^i,
\label{mom6}
\\
&&S=\frac{1}{2}\sum_{ni}\int dtp_{ni}^\ast p_{ni}. \label{act6}
\end{eqnarray}
For the translation mode in particular we have
$du_{0i}/dt=p_{0i}$, $dp_{0i}/dt=0$, yielding $u_{0i}=p_{0i}t$ and
we obtain
\begin{eqnarray}
u=u_{\text{dw}}+p_{0i}\sigma_i
t\left(\frac{3k_0}{4}\right)^{1/2}k_0^{-1} \nabla u_{\text{dw}}
\propto u_{\text{dw}}(x-v_it), \label{dwu4}
\end{eqnarray}
with propagation velocity
\begin{eqnarray}
v_i=-p_0\sigma_i\left(\frac{3}{4k_0}\right)^{1/2}. \label{vel2}
\end{eqnarray}
Likewise, the dynamics of the domain wall at $x_i$ is given by
\begin{eqnarray}
&&E_{0i}=\frac{1}{2}p_{0i}^2, \label{e3}
\\
&&\Pi_{0i}=-p_{0i}\sigma_i\left(\frac{4k_0}{3}\right)^{1/2},
\label{m3}
\\
&&S_0=\frac{1}{2}Tp_{0i}^2. \label{s3}
\end{eqnarray}
Summarizing, the linear analysis of the static domain wall
configuration leads to a picture of a dilute gas of propagating
domain walls. Superposed on the domain walls are localized
deformation modes and extended modes of diffusive character. The
time evolution of the domain wall-linear mode gas is moreover
subject to three constraints: 1) the topological signature of the
domain walls implies that a right hand domain wall is matched to a
consecutive left hand domain wall, 2) translational invariance
implies that the total momentum $\Pi$ given by Eq. (\ref{mom6}) is
conserved, and 3) finally, time translation invariance entails
that the total energy $E$ given by Eq. (\ref{ham6}) is a constant
of motion.
\subsection{Domain wall nucleation and annihilation}
\begin{figure}
\includegraphics[width=.5\hsize]{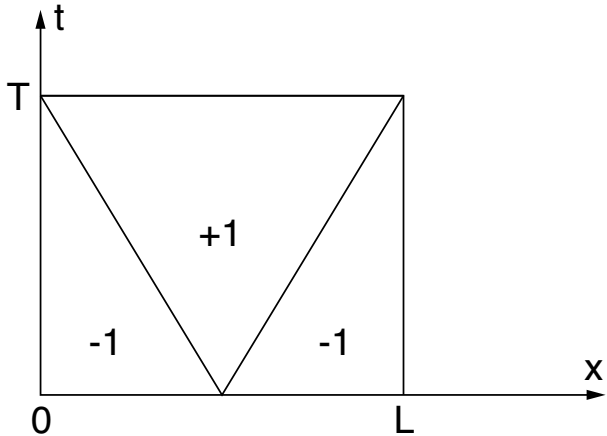}
\caption{ We show the transition from the ground state $u=-1$ to
the ground state $u=+1$ in a system of size $L$ in time $T$ due to
the nucleation of a domain wall pair at the center of the system.}
\label{fig5}
\end{figure}

\begin{figure}
\includegraphics[width=0.5\hsize]{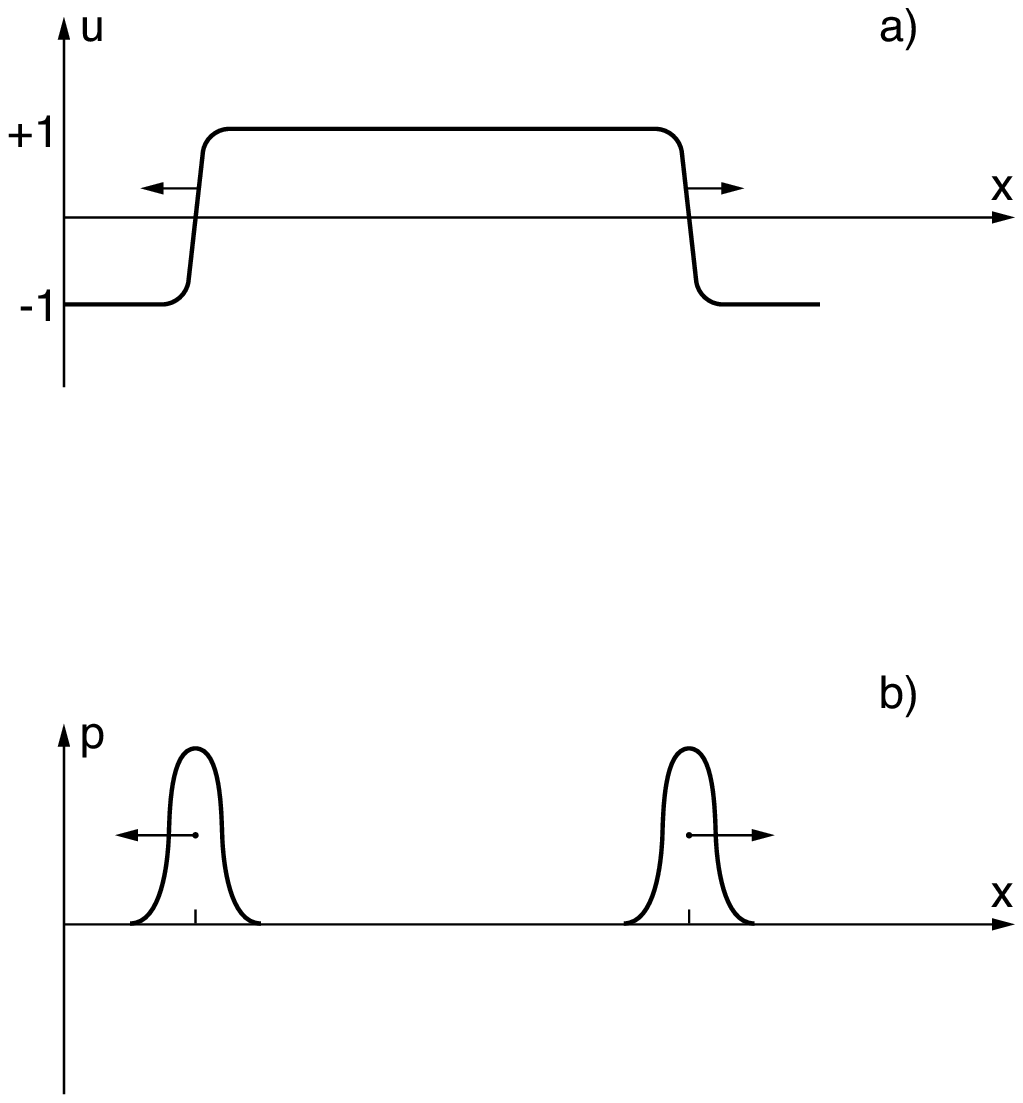}
\caption{In a) we show the domain wall pair propagating away from
the nucleation zone. In b) we show the associated comoving noise
field profiles.}\label{fig6}
\end{figure}

The above analysis of the dynamics of the domain wall - diffusive
mode system is restricted to the dilute gas regime. However, the
topological constraints together with the conservation laws allow
a heuristic analysis of domain wall collisions. When a right hand
domain wall collides with a left hand domain wall the topological
constraint implies that they must annihilate to the uniform ground
states $u=+1$ or $u=-1$. Moreover, since Eq. (\ref{mom1}) implies
that the momentum $\Pi=0$ for $u=\pm 1$, the domain wall pair
prior to collision must move with equal and opposite momenta,
i.e., equal and opposite velocities. Since the phase space
formulation is time reversal invariant we also infer that pairs of
oppositely moving domain walls of opposite parity can nucleate out
of the uniform ground states $u=\pm 1$. In the ground state Eq.
(\ref{ham2}) implies that the energy is given by $E=(1/2)\int
dx~p^2$. Consequently, in order to generate domain wall pairs out
of the ground state we must assign a finite noise field $p$. In
Fig. ~\ref{fig5} we have in a plot of $t$ versus $x$ depicted the
transition from the state $u=-1$ to the state $u=+1$ due to the
formation of a domain wall pair. The system is of size $L$ and the
transition takes place in time $T$.
In Fig. ~\ref{fig6} we have shown the corresponding propagation of
the domain wall pair receeding from the nucleation zone with
opposite velocities together with the associated noise field.
According to Eqs. (\ref{vel1}) and (\ref{solp}) the noise field
profiles comoving with the domain walls are positive and have the
form $1/\cosh^2k_0(x-vt)$
%
\section{\label{stochastic} Stochastic interpretation}
The domain wall gas picture introduced above allow a systematic
dynamical approach to the determination of kinetic pathways and to
the evaluation of the Arrhenius factors associated with the
transitions. The approach moreover permits a straightforward
stochastic interpretation making contact with the customary
discussion of the noisy Ginzburg-Landau equation.
\subsection{Domain wall random walk}
\begin{figure}
\includegraphics[width=0.5\hsize]{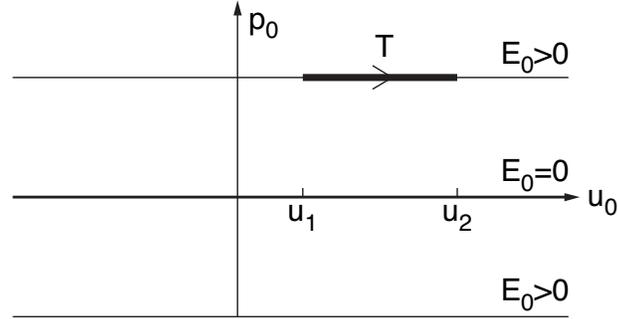}
\caption{ We show the orbits in the $(u_0,p_0)$ phase space in the
case of domain wall motion. The finite-time orbits lie on the
finite energy manifolds $E_0=(1/2)p_0^2$. In the long time limit
$T\rightarrow\infty$ the orbit from $u_1$ to $u_2$ migrates to the
zero-energy manifold $E_0=0$. } \label{fig7}
\end{figure}

In the case of a single domain wall the dynamics follows from Eqs.
(\ref{linu3}) and (\ref{linp3}) for $\lambda_0=0$, i.e., the
equations of motion $du_0/dt=p_0$ and $dp_0/dt=0$ with solutions
$u_0=p_0t$, $p_0=\text{constant}$. The energy $E_0$, momentum
$\Pi_0$ and action $S_0$ are given by (\ref{e2}), (\ref{m2}), and
(\ref{s2}), respectively, and the orbits lie on the energy
manifolds $E_0=(1/2)p_0^2$. In Fig.~\ref{fig7} we have depicted
the phase space for the motion of a single domain wall. The plot
depicts orbits from $u_1$ to $u_2$ in time $T$ on the manifold
$E_0=(1/2)p_0^2$. In the long time limit the orbits migrate to the
zero-energy manifold $E=0$. The phase space plot is a degenerate
limit of the phase space plot in Fig.~\ref{fig3} for
$k_0\rightarrow 0$. We note the absence of a stationary
zero-energy manifold and saddle point yielding a stationary state.
Since the momentum $\Pi_0$ according to Eq. (\ref{m2}) is
proportional to the canonical momentum $p_0$ it follows directly
from the canonical structure that the conjugate variable to
$\Pi_0$ is the center of mass position $x_0$ of the domain wall.
From the Poisson bracket $\{u_0,p_0\}=1$ and $\{x_0,\Pi_0\}=1$
together with $\Pi_0=-\sigma(4k_0/3)^{1/2}p_0$, we infer
$u_0=-\sigma m^{1/2}x_0$, $m=4k_0/3$, and the ensuing equation of
motion
\begin{eqnarray}
m\frac{dx_0}{dt}=\Pi_0, \label{eqx1}
\end{eqnarray}
with solution
\begin{eqnarray}
x_0=\frac{\Pi_0}{m} t ~+~\text{constant}. \label{solx1}
\end{eqnarray}
Likewise, the energy and action are given by $E_0=\Pi_0^2/2m$ and
\begin{eqnarray}
S_0 = T\frac{\Pi_0^2}{2m}. \label{act7}
\end{eqnarray}
The stochastic interpretation of the phase space dynamics readily
follows. Considering an orbit from $x_1$ to $x_2$ traversed in
time $T$, corresponding to the propagation of a domain wall with
center of mass position $x_1$ at time $t=0$ to the center of mass
position $x_2$ at time $t=T$, and inserting the solution
(\ref{solx1}) in Eq. (\ref{s2}) we obtain
\begin{eqnarray}
S_0 = \frac{1}{2}m\frac{(x_2-x_1)^2}{T}, \label{act8}
\end{eqnarray}
yielding the transition probability from $x_1$ to $x_2$ in the
time interval $T$
\begin{eqnarray}
P(x_1,x_2,T)\propto\exp
\left[-\frac{m}{2\Delta}\frac{(x_2-x_1)^2}{T}\right]. \label{dis3}
\end{eqnarray}
This is the Gaussian distribution for random walk with root mean
square deviation $x_{\text{rms}}=(2T\Delta/m)^{1/2}$, $m=4k_0/3$.
We note that the distribution in Eq. (\ref{dis3}) is obtained in
the limit $\Gamma\rightarrow 0$ from the overdamped oscillator
distribution in Eq. (\ref{dis2}). We conclude that the uniform or
ballistic motion of a domain wall within the dynamical description
corresponds precisely to ordinary random walk of the domain wall
within the associated stochastic description.

The random walk behavior of a domain wall also follows easily from
the Ginzburg Landau equation (\ref{gl1}). Inserting the
fluctuating domain wall ansatz $u(x,t)=\tanh k_0(x-x(t))$, where
$x(t)$ is the time-dependent center of mass $x$ and noting that
$\delta F/\delta u=0$ for a domain wall we obtain, integrating
over space, setting $\eta(t)=\int dx~\eta(x,t)$,
\begin{eqnarray}
\frac{dx(t)}{dt}\propto\eta(t), \label{lanrw}
\end{eqnarray}
which is the Langevin equation for random walk.

It is also straightforward to include the contribution to the
stochastic behavior from the deformation and diffusive modes
associated with the domain wall. From Eq. (\ref{act3}) applied to
the local deformation mode (def) $u_1$ and the diffusive modes
(diff) $u_k$ we obtain for the total action for a dressed domain
wall
\begin{eqnarray}
S=S_0+S_{\text{def}}+S_{\text{diff}}, \label{act9}
\end{eqnarray}
where $S_0$ is given by Eq. (\ref{act8}) and $S_{\text{dm}}$ and
$S_{\text{b}}$ by
\begin{eqnarray}
&&S_{\text{def}}= 3\Gamma k_0^2\frac{(u_{21}-u_{11}e^{-3\Gamma
k_0^2T})^2} {1-e^{-6\Gamma k_0^2T}}, \label{sdm}
\\
&&S_{\text{diff}}=\int\frac{dk}{2\pi} \Gamma(k^2+4k_0^2)
\frac{|u_{2k}-u_{1k}e^{-\Gamma(k^2+4k_0^2)T}|^2}
{1-e^{-2\Gamma(k^2+4k_0^2)T}}. \label{sb}
\end{eqnarray}
For the transition probability from an initial dressed domain wall
configuration $u_1(x)=\{x_1, u_{11},u_{1k}\}$ to a final dressed
domain wall configuration $u_2(x)=\{x_2,u_{21},u_{2k}\}$ during
the time interval $T$ we finally obtain
\begin{eqnarray}
P(u_1,u_2,T)=P_0P_{\text{def}}P_{\text{diff}}, \label{distot1}
\end{eqnarray}
where $P_0$ is given by Eq. (\ref{dis3}) and
$P_{\text{def}}\propto\exp[-S_{\text{def}}/\Delta]$ and
$P_{\text{diff}}\propto\exp[-S_{\text{diff}}/\Delta]$. In the long
time limit $t\rightarrow\infty$ the domain wall performs a random
walk, whereas the deformation mode and diffusive modes attain a
stationary state, i.e.,
\begin{eqnarray}
P(x,T;u_1,\{u_k)\})\propto\exp\left[-\frac{m x^2}{2\Delta
T}\right]\exp \left[-\frac{3\Gamma k_0^2u_1^2}{\Delta}\right]\exp
\left[-\frac{1}{\Delta}
\int\frac{dk}{2\pi}\Gamma(k^2+4k_0^2)|u_k|^2 \right].
\label{distot2}
\end{eqnarray}
%
\subsection{Domain wall gas}
In the case of a dilute domain wall gas with associated linear
modes the discussion above applies in a generalized form. From
Eqs. (\ref{linu5}) and (\ref{linp5}) for $\lambda_0=0$ we obtain
the equations of motion $du_{0i}/dt=p_{0i}$ and $dp_{0i}/dt=0$
with solutions $u_{0i}=p_{0i}t+\text{cst}$, $p_{0i}=\text{cst}$.
The energy, momentum, and action are given by Eqs. (\ref{e3}),
(\ref{m3}), and (\ref{s3}), respectively, and each domain wall
lies on the corresponding energy manifold $E_{0i}=(1/2)p_{0i}^2$.
For each domain wall the phase space plot in Fig.~\ref{fig4} thus
applies. Since the action in (\ref{act6}) is additive with a
contribution from each domain wall the transition probabilities
factorize and we obtain for the random walk part from (\ref{dis3})
\begin{eqnarray}
P(\{x_{i1}\},\{x_{i2}\},T) \propto
\prod_i\exp\left[-\frac{m(x_{i2}-x_{i1})^2}{2\Delta T} \right].
\label{dis4}
\end{eqnarray}
Likewise, the contributions from the deformation and diffusive
modes follow from Eqs. (\ref{sdm}) and (\ref{sb}).
\section{Kinetic transitions}
In the previous sections we have established the dynamical
framework in the case of the noise-driven Ginzburg-Landau equation
and established the connection to the customary  stochastic
interpretation. In summary, the domain wall gas picture provides a
description of a switching scenario in terms of moving right hand
and left hand domain walls with associated linear modes. The
dynamical approach moreover implies that the kinetic pathway from
an initial configuration to a final configuration is associated
with a specific orbit in canonical phase space. The action
associated with the orbit yields the Arrhenius factor associated
with the transition. The domain wall gas picture shows that a
class of orbits, i.e., a class of solutions of the field equations
can be parametrized in terms of a dilute gas of propagating domain
walls with superimposed local deformation and extended
phase-shifted diffusive modes. This picture is derived from a
linear analysis and only holds a priori in a dilute gas limit and
at short times. As discussed below the domain wall dynamics at
later times can be extracted heuristically from a combination of
selection rules and conservation laws.
\subsection{Equilibrium}
\begin{figure}
\includegraphics[width=0.5\hsize]{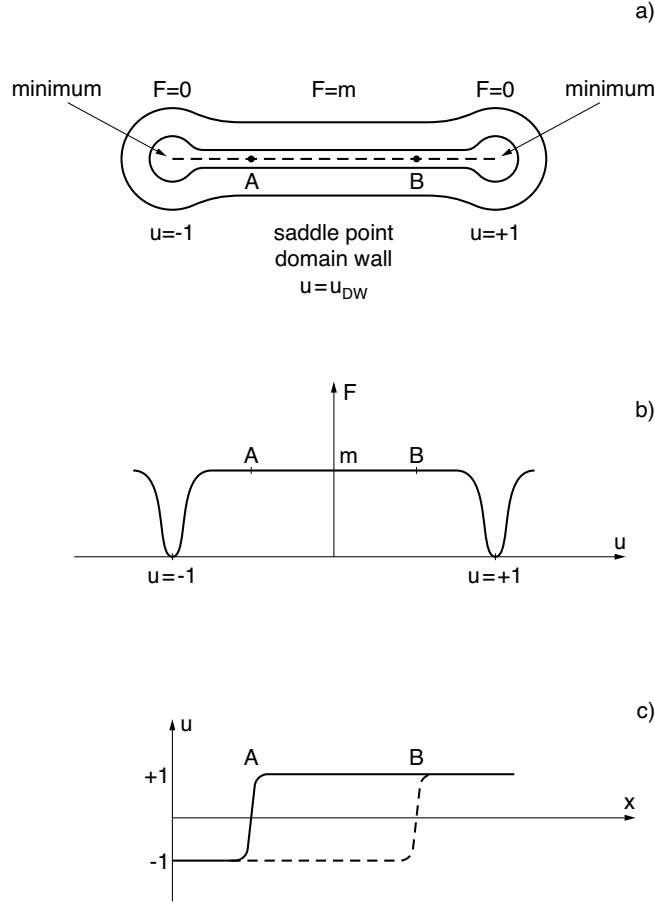}
\caption{In a) we have in a contour plot sketched the free energy
landscape in the case of a single static domain wall overlapping
for large $|x|$ with the ground states $u=-1$ and $u=+1$. The
points A and B indicate two distinct positions of the domain wall,
corresponding to the translation mode. In b) we have sketched the
free energy as function of $u$. The $u$ axis labels collectively
the $u(x)$ configurations in the free energy landscape. The points
A and B refer to the domain wall positions. Since the free energy
of the domain wall is independent of the center of mass position
the landscape exhibits a constant $F$ ridge. In c) we have
depicted the corresponding domain wall configurations at center of
mass positions A and B.} \label{fig8}
\end{figure}

\begin{figure}
\includegraphics[width=0.5\hsize]{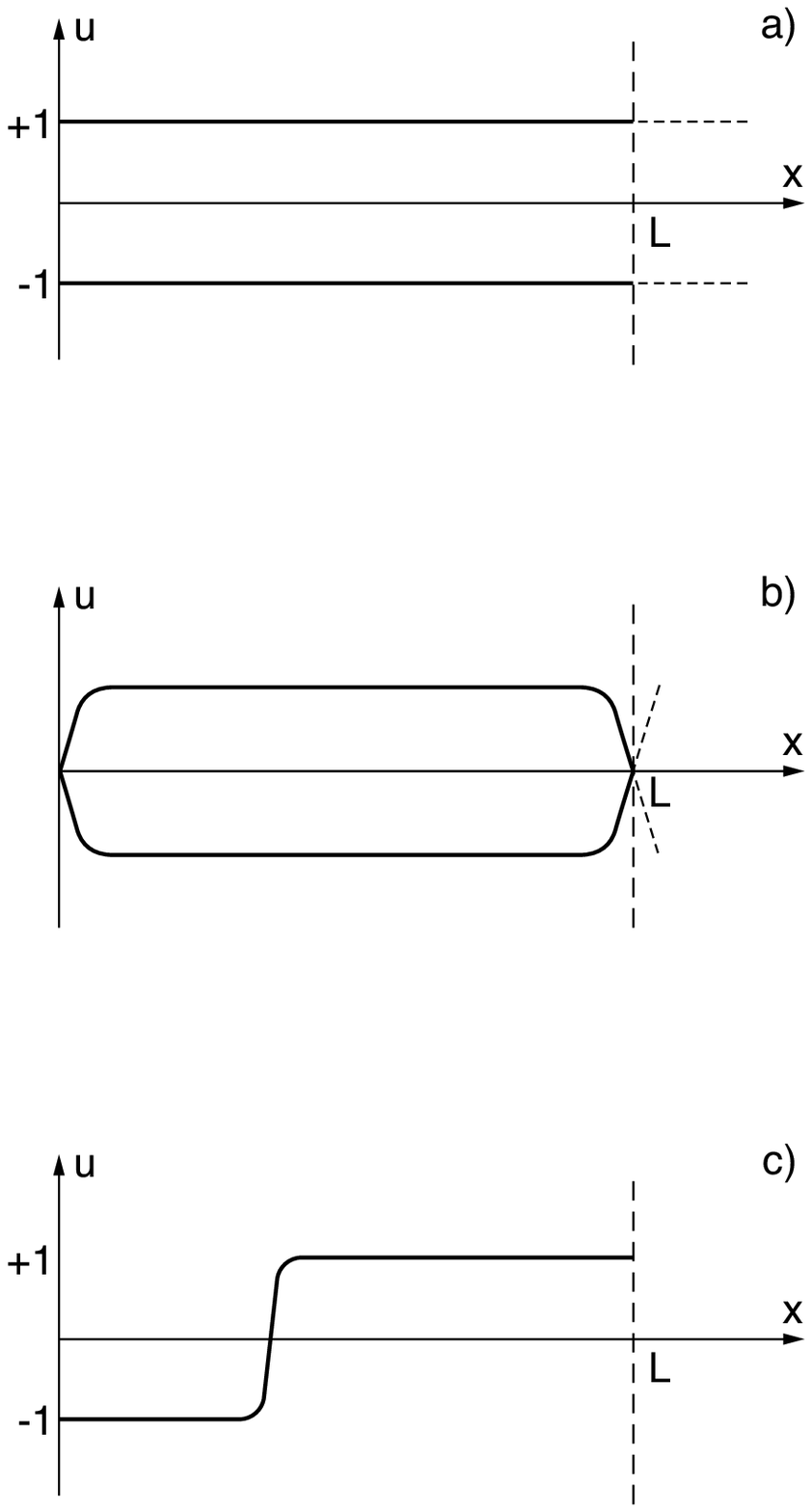}
\caption{ In a) we have shown the ground state configurations with
$F=0$ in the case of open or periodic boundary conditions. In b)
the lowest configuration with fixed boundary conditions have
$F=m$. In c) we show the lowest spatially degenerate domain wall
configuration with $F=m$ in the case of skew boundary conditions.
} \label{fig9}
\end{figure}

\begin{figure}
\includegraphics[width=0.5\hsize]{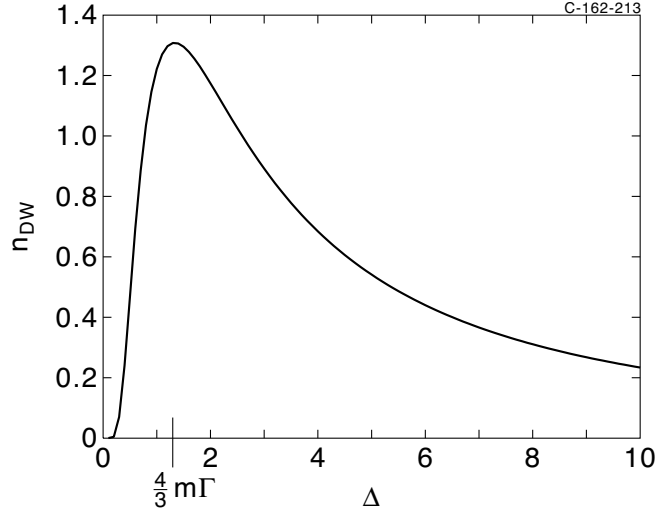}
\caption{We show the domain wall density $n_{\text{dw}}$ in units
of $\Delta/2\Gamma$ as function of the noise strength $\Delta$ in
units of $\Gamma m$. The density vanishes in the limits
$\Delta\rightarrow 0$ and $\Delta\rightarrow\infty$ and exhibits a
maximum at $4m\Gamma/3$.} \label{fig10}
\end{figure}

Before we discuss kinetic transitions it is instructive first
briefly to discuss the equilibrium properties. In equilibrium the
free energy landscape is determined by the structure of $F(u)$ in
Eq. (\ref{free}). The complex landscape is characterized by the
presence of two global minima corresponding to the uniform ground
states $u=+1$ and $u=-1$ and an infinity of saddle points
corresponding to static nonuniform multi-domain wall
configurations. The flat parts of the saddle points correspond to
the translation modes associated with the domain walls. The free
energy $F=0$ for the ground states. For an n-domain wall
configuration $F=nm$, where the mass $m=4k_0/3$. In
Fig.~\ref{fig8} we have in sketched the free energy landscape
associated with a single static domain wall connecting $u=-1$ and
$u=+1$.
In the case of zero noise for $\Delta=0$ the system starting from
an arbitrary initial configuration $u_1(x)$ approaches the state
with lowest free energy compatible with the imposed boundary
conditions. The relaxational dynamics is governed by Eq.
(\ref{gl2}) and corresponds in the dynamical approach to an orbit
confined to the $p=0$ zero-energy submanifold. For open or
periodic boundary conditions the configurations with lowest free
energy are the two degenerate uniform ground states $u=\pm 1$.
Imposing vanishing boundary conditions $u=0$ at $x=0$ and $x=L$
for a system of size $L$ the field $u$ grows to the uniform values
$u=-1$ or $u=+1$ over a healing length of order $1/k_0$. Imposing
periodic boundary conditions we infer that the healing profile
corresponds to a half domain wall and that the free energy of the
configuration equals $2\times(1/2)m=m$. Finally, for skew boundary
conditions with $u=-1$ for $x=0$ and $u=+1$ for $x=L$ the state
with lowest free energy corresponds to a single domain wall with
free energy $m$. The different scenarios are depicted in
Fig.~\ref{fig9}.
In the presence of noise for $\Delta\neq 0$ the ground states
become metastable and the system populates the excited states with
$F\neq 0$. The partition function provides a global
characterization of the equilibrium excursions in the free energy
landscape and is according to Eq. (\ref{stat1}) given by
\begin{eqnarray}
Z=\sum_{\{u\}}\exp\left[-\frac{2\Gamma}{\Delta}F(u)\right],
\label{par1}
\end{eqnarray}
where the configuration $\{u\}$ has the free energy $F(u)$,
yielding the Boltzmann factor in Eq. (\ref{par1}). The evaluation
of $Z$ and associated correlations, e.g., $\langle uu\rangle(x,t)$
in the dilute non-overlapping domain wall gas limit follows
closely the well-known soliton and instanton methods developed in
the seventies and used in Ref. \cite{Fogedby85} for the classical
easy-plane ferromagnet. Expanding the free energy defined by Eqs.
(\ref{free}) and (\ref{pot}) about an n-domain wall configuration
we have, setting $u=u_{\text{dw}}+\delta u$,
$F=F(u_{\text{dw}})+(1/2)\delta uD\delta u$. Using
$F(u_{\text{dw}})=nm$, inserting $D$ from Eq. (\ref{op2}), and
expanding $\delta u$ on the eigenfunctions $\Psi_n$, we obtain
\begin{eqnarray}
F=nm+\frac{1}{2}\sum_{ni}\lambda_n|u_{ni}|^2.
\end{eqnarray}
Introducing a domain wall chemical potential $\mu_{\text{dw}}$ in
order to control the domain wall densities we arrive at the grand
partition function
\begin{eqnarray}
Z=2\sum_{n=0}^\infty\exp
\left[\frac{2\Gamma}{\Delta}n(\mu_{\text{dw}}-m) \right]
\frac{1}{n!}\int\prod_{i=1}^ndu_{0i}du_{1i}\prod_kdu_{ki}
\exp\left[-\frac{\Gamma}{\Delta}\left(\lambda_0\sum_{i=1}^nu_{0i}^2
+\lambda_1\sum_{i=1}^nu_{1i}^2 + \lambda_k\sum_k|u_{ki}|^2\right)
\right]. \label{par2}
\end{eqnarray}
The overall factor $2$ arises from the double degeneracy of the
ground state and the domain walls connecting the ground states.
The $\lambda_0=0$ eigenvalue yields the translation mode and it
follows that $du_{0i}=m^{1/2}dx_i$, where $x_i$ is the position of
the i-th domain wall. The factor $1/n!$ takes into account the
ordering of the domain wall when integrating $x_i$ over a system
of size $L$. Performing the Gaussian integrals over the
deformation and diffusive modes we have in more reduced form
\begin{eqnarray}
Z=2\sum_{n=0}^\infty\frac{1}{n!}\exp
\left[\frac{2\Gamma}{\Delta}n(\mu_{\text{dw}}-m) \right]
(m\Lambda)^{n/2}L^n
\left(\frac{\pi\Delta\Lambda}{\Gamma\lambda_1}\right)^{n/2}
\exp\left[\frac{1}{2}\sum_k\log\frac{\pi\Delta\Lambda}
{\Gamma\lambda_k}\right]. \label{par3}
\end{eqnarray}
We have introduced the large wavenumber UV cutoff $\Lambda$ in
order to regularize $Z$ at small distances. Note that in the
treatment in Ref. \cite{Fogedby85} the lattice distance of the
magnetic chain provided a natural small scale cutoff. Replacing
the summation over $k$ by an integral according to the
prescription $\sum_k\rightarrow\int dk\rho_k$, where the density
of states $\rho_k=L/2\pi+(n/2\pi)d\delta/dk$,
$\delta=2\tan^{-1}(k_0/k)+ 2\tan^{-1}(2k_0/k)$, and summing over
domain walls the partition function factorizes into a diffusive
part and a domain wall part, incorporating also the contribution
from the localized deformation modes,
\begin{eqnarray}
Z&=&Z_{\text{diff}}Z_{\text{dw}}, \label{par4}
\\
Z_{\text{diff}}&=& \exp\left[\frac{L\Lambda}{2\pi}
\log\frac{\pi\Delta}{\Gamma\Lambda}\right],\label{par5}
\\
Z_{\text{dw}}&=&2\exp\left[8Lk_0^{3/2}
\left(\frac{\Gamma}{\pi\Delta}\right)^{1/2}
\exp\left(\frac{2\Gamma}{\Delta}(\mu_{\text{dw}}-m)\right)\right].
\label{par6}
\end{eqnarray}
In a thermal environment $\Delta=2\Gamma T$ and the partition
function gives direct access to e.g., the specific heat. We shall
not pursue this calculation here except noting that the gap in the
domain wall excitation spectrum gives rise to a Schottky anomaly
in the specific heat, see e.g., Ref. \cite{Fogedby85}. Noting that
the number of domain walls accessed by the stationary fluctuations
is undetermined it is, however, instructive to evaluate the domain
wall density $n_{\text{dw}}$. From the structure of $Z$ we infer
$n_{\text{dw}}=(\Delta/2\Gamma)(d\log Z/d\mu)_{\mu=0}$ and
inserting Eqs. (\ref{par4}), (\ref{par5}), and (\ref{par6}) we
obtain, setting $k_0=3m/4$,$x\to f$
\begin{eqnarray}
n_{\text{dw}}=8\pi^{-1/2}\left(\frac{\Delta}{\Gamma}\right)
\left(\frac{3\Gamma}{4\Delta}m\right)^{3/2}
\exp\left(-\frac{2\Gamma}{\Delta}m\right).
\end{eqnarray}
The domain wall density vanishes in the limits $\Delta\rightarrow
0$ and $\Delta\rightarrow\infty$ and exhibits a maximum for
$\Delta\approx\Gamma m$. We have depicted $n_{\text{dw}}$ as a
function of $\Delta$ in Fig.~\ref{fig10}.
%
\subsection{Transitions in the Kramers case}
%
Before turning to the noise-induced kinetic transitions in the
Ginzburg-Landau equation it is instructive to review the classical
Kramers theory \cite{Kramers40,Haenggi90} for a single degree of
freedom, which can be analyzed completely, and its formulation
within the canonical phase space approach.
%
\subsubsection{Kramers theory}
%
We consider the overdamped case characterized by the Langevin
equation for one degree of freedom $u$
\begin{eqnarray}
&&\frac{du}{dt}=-\Gamma\frac{dF}{du} + \eta, \label{lan2}
\\
&&\langle\eta(t)\eta(0)\rangle = \Delta\delta(t), \label{noise2}
\end{eqnarray}
with kinetic coefficient $\Gamma$ and noise strength $\Delta$; in
thermal equilibrium $\Delta = 2\Gamma T$, where $T$ is the
temperature. For the free energy $F$ we assume a double well
profile with maximum $F(0)$ at $u=0$ and minima at $u=\pm 1$ with
$F(\pm 1)=0$. The free energy is depicted in Fig.~{\ref{fig11} a).
\begin{figure}
\includegraphics[width=0.5\hsize]{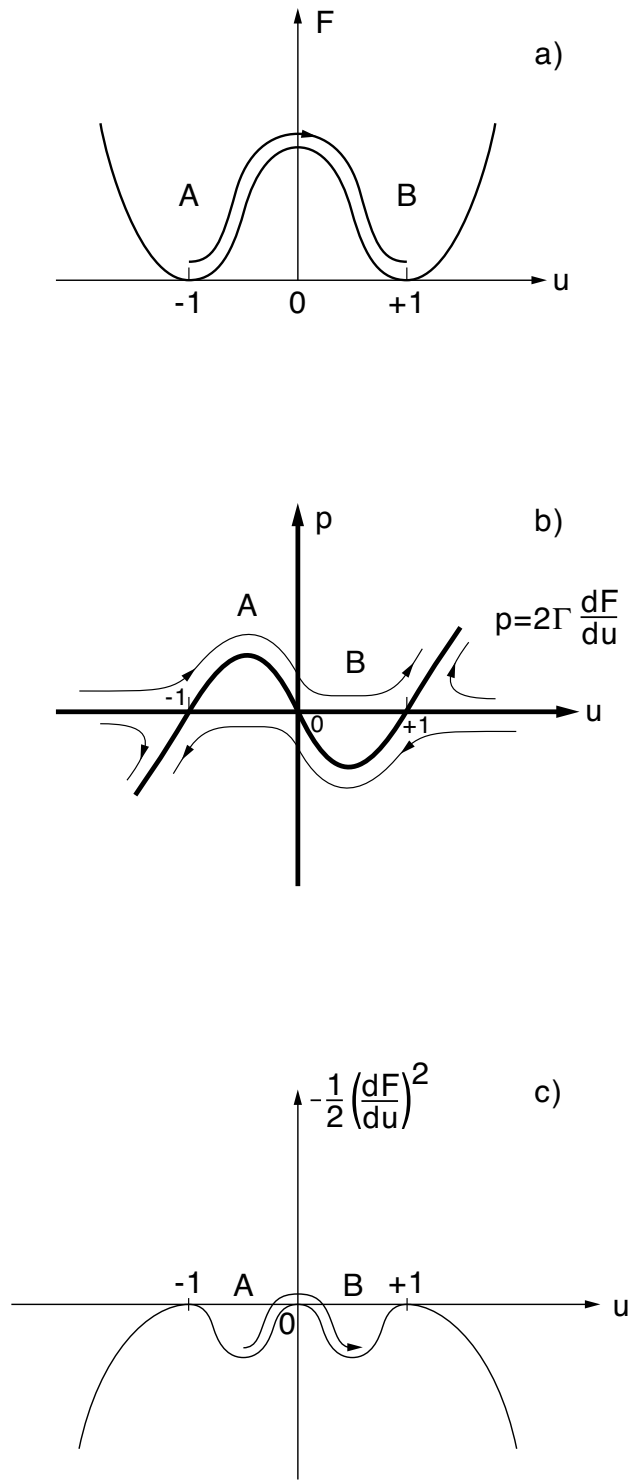}
\caption{In a) we show the double well structure of the free
energy $F$. In b) we show orbits in the canonical phase space. The
zero-energy manifolds are $p=0$ and $p=2\Gamma dF/du$,
intersecting at the saddle points $(u,p)=(-1,0)$, $(0,0)$, and
$(1,0)$. In c) we show the inverted double well potential entering
in the Newton equation description of the transition.}
\label{fig11}
\end{figure}
The Fokker-Planck equation associated with Eqs. (\ref{lan2}) and
(\ref{noise2}) has the form
\begin{eqnarray}
\frac{\partial P}{\partial
t}=\frac{1}{2}\Delta\frac{\partial^2P}{\partial
u^2}+\frac{\partial}{\partial u}\left(\Gamma\frac{dF}{du}P\right).
\label{fp2}
\end{eqnarray}
We note the stationary state for $\partial P/\partial t=0$
\begin{eqnarray}
P_{\text{st}}\simeq\exp\left[-\frac{2\Gamma}{\Delta}F\right],
\label{stat3}
\end{eqnarray}
in accordance with Eq. (\ref{stat1}), showing how the free energy
profile is globally sampled in the stationary state.

In order to evaluate the transition rate from, say, the state
$u=-1$ to the state $u=+1$ across the free energy barrier $F(0)$
we follow Kramers and set up a constant probability current $J$
across the barrier generated by a source to the left of the
barrier in region $A$ and absorbed by a sink to the right of the
barrier in region $B$. From the conservation law $\partial
P/\partial t+\partial J/\partial u=0$ and Eq. (\ref{fp2}) we
derive the current
\begin{eqnarray}
J=-\frac{1}{2}\Delta\frac{\partial P}{\partial u} -
\Gamma\frac{dF}{du}P. \label{cur}
\end{eqnarray}
Assuming a sink at $u_+>0$ we obtain the steady state solution
\begin{eqnarray}
P(u)=\frac{2J}{\Delta}\exp\left({-\frac{2\Gamma}{\Delta}F(u)}\right)
\int_u^{u_+}du'\exp\left({\frac{2\Gamma}{\Delta}F(u')}\right).
\end{eqnarray}
Setting up a population $n=\int_{-\infty}^0 du P(u)$ to the left
of the barrier the rate is given by $k=J/n$ and we have
\begin{eqnarray}
k^{-1}=\frac{2}{\Delta}\int_{-\infty}^0du
\exp\left({-\frac{2\Gamma}{\Delta}F(u)}\right)
\int_u^{u_+}du'\exp\left({\frac{2\Gamma}{\Delta}F(u')}\right).
\end{eqnarray}
Finally, a simple steepest descent calculation for
$\Delta\rightarrow 0$ yields Kramers celebrated result for the
rate in the overdamped Smoluchowski limit, $F''=d^2F/du^2$,
\begin{eqnarray}
k=\frac{\Gamma}{2\pi}(F''(-1)|F''(0)|)^{1/2}
\exp\left({-\frac{2\Gamma}{\Delta}F(0)}\right). \label{rate}
\end{eqnarray}
The rate is determined by the Arrhenius factor
$\exp(-(2\Gamma/\Delta)F(0))$ depending on the height of the
barrier and the prefactor $(\Gamma/2\pi)(F''(-1)|F''(0)|)^{1/2}$.
Here the double derivative $F''$ can be associated with the
oscillation frequencies in the potential well at $u=-1$ and about
the maximum at $u=0$
%
\subsubsection{Dynamical interpretation of Kramers theory}
%
Here we discuss the Kramers escape problem from a potential well
within the canonical phase space method. According to the general
formulation in Sec. III the Hamiltonian associated with the
Langevin equation (\ref{lan2} has the form
\begin{eqnarray}
H=\frac{1}{2}p\left(p-2\Gamma\frac{dF}{du}\right),
\end{eqnarray}
yielding the equations of motion
\begin{eqnarray}
&&\frac{du}{dt}=-\Gamma\frac{dF}{du}+p, \label{eq1k}
\\
&&\frac{dp}{dt}=\Gamma\frac{d^2F}{du^2}p, \label{eq2k}
\end{eqnarray}
and the associated action
\begin{eqnarray}
S(u,T)=\int^{u,T}dt\left[p\frac{du}{dt}-H\right]. \label{actk}
\end{eqnarray}
For the normalized transition probability we have
\begin{eqnarray}
P(u,T)=\frac{\exp\left[-\frac{S(u,T)}{\Delta}\right]} {\int
du\exp\left[-\frac{S(u,T)}{\Delta}\right]}
\end{eqnarray}
In Fig.~\ref{fig11} b) we have depicted the phase space for the
Kramers escape case. The zero-energy manifolds are given by $p=0$
and $p=2\Gamma dF/du$. On the phase space plot we have shown an
orbit from region $A$ to region $B$, i.e., across the free energy
barrier. Comparing Fig.~\ref{fig11} a) with Fig.~\ref{fig11} b) we
note that the ``uphill'' part of the orbit in region $A$ is
controlled by the $p=2\Gamma dF/du$ manifold whereas the
``downhill'' part in region $B$ is controlled by the $p=0$
manifold. In the long time limit the orbit approaches the
zero-energy manifolds $p=0$ and $p=2\Gamma dF/du$ passing close to
the saddle points at $u=-1$, $u=0$, and $u=1$. A simple
calculation along the ``bulge'' setting $H=0$ and inserting
$p=2\Gamma dF/du$ in Eq. (\ref{actk}) yields the Arrhenius factor
\begin{eqnarray}
\exp\left[-\frac{2\Gamma}{\Delta}F(0)\right],
\end{eqnarray}
in accordance with Kramers result in Eq. (\ref{rate}).

Finally, eliminating $p$ in the coupled equations (\ref{eq1k}) and
(\ref{eq2k}) we obtain the Newton equation of motion
\begin{eqnarray}
\frac{d^2u}{dt^2}=-\Gamma^2\frac{d}{du}
\left[-\frac{1}{2}\left(\frac{dF}{du}\right)^2\right],
\end{eqnarray}
for the motion of a particle of mass $1/\Gamma^2$ in the potential
$-(1/2)(dF/du)^2$. In Fig.~\ref{fig11} c) we have shown the
potential which for a double well free energy possesses three
maxima at $u=0$ and $u=\pm 1$. As indicated in Fig.~\ref{fig11} c)
the long time orbit from $A$ to $B$ is associated with the long
waiting time at the maximum for $u=0$

\subsection{Transitions in the Ginzburg-Landau case}
In order to induce a non-equilibrium kinetic transition across the
free energy landscape in the Ginzburg-Landau case  we fix the
initial configuration $u_1(x)$ at time $t=0$ and the final
configuration $u_2(x)$ at time $T$. There are two fundamental
issues: 1) the determination of the kinetic pathways and 2) the
evaluation of the transition rate. Generally, in order to minimize
the free-energy cost, the pathway passes via saddle points, i.e.,
the multi-domain wall excitations, in the free energy landscape.
The actual path chosen, however, depends on the time $T$ allocated
to the transition. The transition rate is determined by the
Arrhenius factor $\exp(-2\Gamma F/\Delta)$ associated with the
free energy $F$ of the saddle points encountered along the
pathway. In addition there is a prefactor determined by the
attempt frequencies; this issue, however, is not dealt with in the
present context.

Focussing on a kinetic transition or switch between the two ground
states $u=-1$ and $u=+1$ the pathway  corresponds to nucleation of
domain walls and their subsequent propagation across the system.
Let us first consider the case of fixed boundary conditions $u=0$
at $x=0$ and $x=L$. The ground state configurations are depicted
in Fig.~\ref{fig9} b). In the presence of noise the ground states
are metastable and the system can switch from $u=-1$ to $u=+1$ by
means of the propagation of a single domain wall across the sample
in the transition time $T$. As discussed above the profiles close
to the boundaries over a healing length of order $k_0^{-1}$
correspond to half domain walls and the free energy of either
ground state is equal to $F_{\text{dw}}=m$, $m=4k_0/3$; note that
for periodic boundary conditions as depicted in Fig.~\ref{fig9} a)
the free energy of the ground states vanishes. In order to
effectuate a transition a domain wall must nucleate at either
boundary and subsequently propagate across the system. The
nucleation free energy is $F_{\text{dw}}=m$ and from our
discussion of the Kramers case we infer the Arrhenius factor
$\exp(-2\Gamma m/\Delta)$ or within the dynamical description a
nucleation action of the order
\begin{eqnarray}
S_{\text{nucl}}=2\Gamma m.
\end{eqnarray}
Within the conventional stochastic description the noise generates
both the nucleation and the subsequent diffusion of the domain
wall across the sample. Since the domain wall carries a finite
energy we must within the dynamical description assign a finite
energy or finite noise field $p(x)$ at $t=0$ in order to ensure
propagation of the domain wall. Considering the case where the
right hand domain wall nucleates at $x=0$, we assign the noise
field $p(x)\propto p_0/\cosh^2 k_0x$ given by Eq. (\ref{solp}) at
time $t=0$. In  terms of the momentum $\Pi_0=-p_0m^{1/2}$ given by
Eq. (\ref{m2}) we thus obtain the velocity $v=\Pi_0/m$, energy
$E_0=\Pi_0^2/2m$, and action by Eq. (\ref{act7}), i.e.,
$S_0=T\Pi_0^2/2m$. The constraint that the system of size $L$ is
switched at time $T$ moreover impose a selection rule on the
propagation velocity. We infer $v=L/T$ and obtain the action
$S_0=mL^2/2T$ associated with the transition. The total action
associated with a single domain wall switch is thus given by
\begin{eqnarray}
S_1(T)=S_{\text{nucl}}+\frac{mL^2}{2T}. \label{as1}
\end{eqnarray}
In Fig.~\ref{fig12} we have in a) depicted the propagation of a
single right hand domain wall across the sample. In b) we have
shown the transition in an $(x,t)$ plot.
\begin{figure}
\includegraphics[width=0.5\hsize]{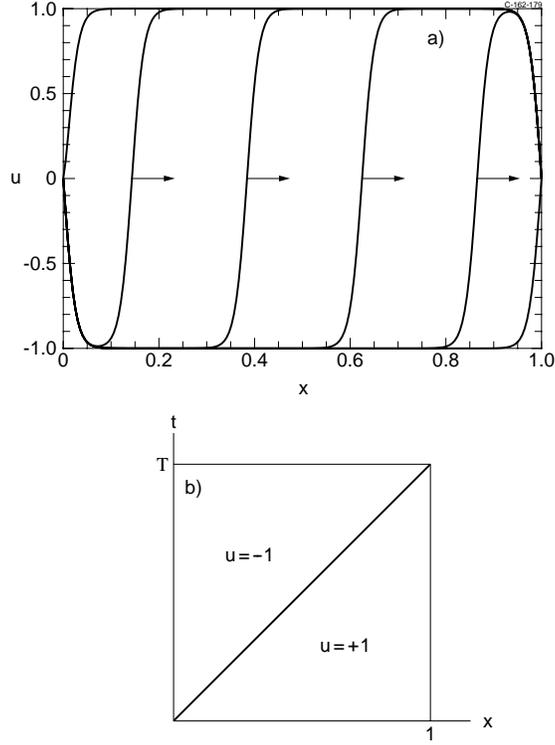}
\caption{In a) the switching from $u=+1$ to $u=-1$ in time $T$ is
effectuated by means of a right hand domain wall propagating with
velocity $v=1/T$. The domain wall is nucleated at $x=0$ and
annihilated at $x=1$. In b) the process is depicted in an $(x,t)$
plot.} \label{fig12}
\end{figure}
In the present case with fixed boundary conditions the momentum
$\Pi_0$ associated with the motion of the domain wall is generated
at the boundary  $x=0$ at time $t=0$ and, subsequently, absorbed
at the boundary $x=L$ at time $t=T$. The momentum is given by Eq.
(\ref{mom3}), i.e.,  $\Pi_0=p_0\int dx u_{\text{dw}}d\Psi_0/dx$.
Inserting $\Psi_0$ and $u_{\text{dw}}$ given by Eqs. (\ref{bs0})
and (\ref{dw1}), respectively, and performing a partial
integration using $\int dx \cosh^{-4}k_0x=4/3k_0$ we obtain
$\Pi_0=-p_0m^{1/2}>0$ since $p_0<0$ for a positive propagation
velocity. Correspondingly, the momentum is absorbed at $x=L$.

For fixed boundary conditions the transition from $u=-1$ to $u=+1$
can also take place by means of nucleation of two domain walls at
both ends of the system. The domain walls subsequently propagate
towards one another and annihilate at $x=L/2$. The scenario is
shown Fig.~\ref{fig13}. Since the action is additive for the
two-domain wall system the nucleation action is given by
$2S_{\text{nucl}}=2m$. However, in order to effectuate the
transition in time $T$ the velocity is half the velocity in the
single domain wall case and we obtain for the action associated
with the 2-domain wall transition
\begin{eqnarray}
S_2(T)=2S_{\text{nucl}}+\frac{mL^2}{4T}. \label{as2}
\end{eqnarray}
\begin{figure}
\includegraphics[width=0.5\hsize]{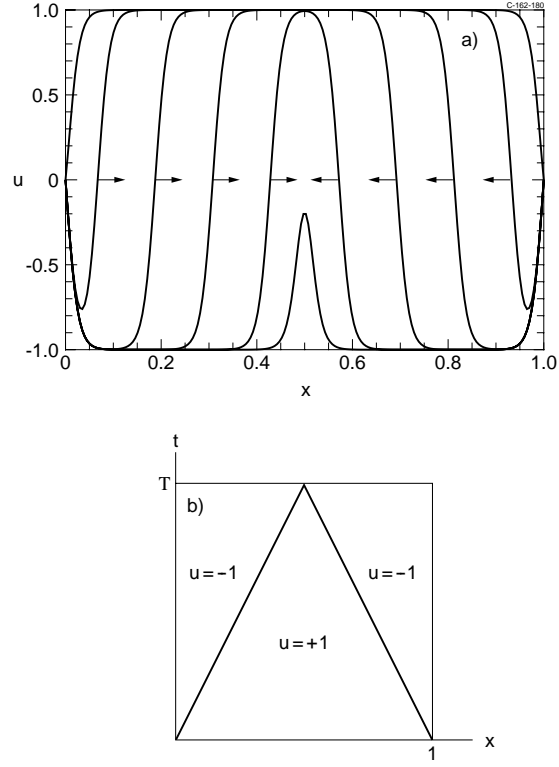}
\caption { In a) the switching from $u=+1$ to $u=-1$ in time $T$
takes place by means of two domain walls propagating in opposite
directions with velocity $v=1/2T$. The domain walls nucleate at
the boundaries and annihilate at the center. In b) the switching
process is depicted in an $(x,t)$ plot.} \label{fig13}
\end{figure}

Finally, in the general case of a transition from $u=+1$ to $u=-1$
in time $T$ by means of the nucleation and propagation of $n$
domain walls the nucleation action is $n S_{\text{nucl}}$ and the
domain walls move with velocity $L/Tn$; we obtain the action
\begin{eqnarray}
S_n(T)=nS_{\text{nucl}}+\frac{mL^2}{2nT}.\label{asn}
\end{eqnarray}
In Fig.~\ref{fig14} we have in an $(xt)$ plot shown a 4-domain
wall switch.
\begin{figure}
\includegraphics[width=0.5\hsize]{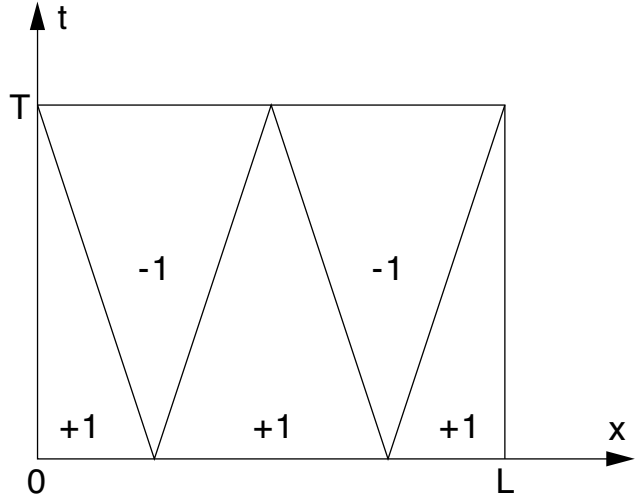}
\caption{We show a 4-domain wall switch in an $(x,t)$ plot.}
\label{fig14}
\end{figure}

In addition to domain wall modes time-dependent localized
deformation and extended diffusive modes are also excited,
corresponding to small Gaussian fluctuations about the local
minima and saddle points and the transition pathway from $u=+1$ to
$u=-1$ proceeds by propagating domain walls with superposed linear
modes subject to energy and momentum conservation and topological
constraints. The energy of the initial state is given by
$E=(1/2)\int dx p^2$ and the noise field thus has to be assigned
initially in order to reach the switched state $u=-1$ in a
prescribed time $T$. For topological reasons the domain walls must
nucleate and annihilate in pairs subject to absorption or
radiation of linear modes, respectively. Since the linear modes
also carry positive action the dynamical modes with lowest action
correspond to nucleation or annihilation of domain wall pairs with
equal and opposite momenta, i.e., equal velocities.

In the case of periodic boundary conditions the momentum $\Pi=\int
dx u\partial p/\partial x$ of the initial and final states is
zero. The system is translational invariant and the formation and
annihilation of one or several domain wall pairs moving with the
same speed take place at equidistant positions along the axis. For
fixed boundary conditions the translational invariance is broken
and the momentum $\Pi$ is nonvanishing corresponding to nucleation
and annihilation of domain walls at the boundaries. This general
scenario of switching is completely consistent with the numerical
analysis in \cite{E02}.
%
\subsubsection{Improved estimate of $S_{\text{nucl}}$}
%
The above estimate of the nucleation action $S_{\text{nucl}}$ for
a domain wall was based on an analogy with the Arrhenius factor in
the simple Kramers case of escape from a potential well. Here we
improve the estimate of $S_{\text{nucl}}$ based on the phase space
formulation. We consider the case of a switch from $u=+1$ to
$u=-1$ in time $T$ proceeding by 1) the nucleation of two domain
walls at the center of the system during the time interval $\delta
t$, 2) propagation of the domain walls with opposite velocities
during the time interval $T-2\delta t$, and 3) the annihilation of
the domain walls at the boundaries during the time interval
$\delta t$. From the width of a single domain wall $k_0^{-1}$ and
the propagation velocity $v=L/2T$ we estimate the nucleation time,
i.e., the time it takes for the nucleation to separate in two
distinct domain walls, to be of order $\delta t\approx
2/vk_0=4T/k_0L$. From Eq. (\ref{act2}) the nucleation action per
domain wall is given by
\begin{eqnarray}
S_{\text{nucl}}=\frac{1}{2}\int_{u=+1,0}^{u,\delta
t}dxdt(pdu/dt-{\cal H}), \label{act10}
\end{eqnarray}
where $u$ is the nucleation configuration just prior to breaking
up into two domain walls; ${\cal H}$ is the energy density which
we determine below. In order to estimate $S_{\text{nucl}}$ we
consider the field equations (\ref{feu2}) and (\ref{fep2}). The
initial configuration is $u(x)=u(x,t=0)=1$ and denoting the
initial noise field by $p_0(x)=p(x,t=0)$ and, moreover,
considering a large system so that $\delta t$ is small, we obtain
to leading order in $\delta t$ from the equations of motion
\begin{eqnarray}
&&u(x,t)\approx\left(\left(\Gamma\frac{\partial^2 u(x,t)}
{\partial x^2}+2\Gamma
k_0^2u(x,t)(1-u(x,t)^2)\right)_{t=0}+p_0(x)\right)\delta t+1,
\\
&&p(x,t)\approx\left(-\Gamma\frac{\partial^2p(x,t}{\partial
x^2}-2\Gamma k_0^2 p(x,t)(1-3u(x,t)^2)\right)_{t=0}\delta
t+p_0(x),
\end{eqnarray}
or by insertion
\begin{eqnarray}
&&u(x,t)\approx p_0(x)\delta t+1,
\\
&&p(x,t)\approx\Gamma\left(-\frac{\partial^2p}{\partial
x^2}+4k_0^2p_0(x)\right)\delta t+p_0(x).
\end{eqnarray}
These solutions describe the initial part of the orbit in $(up)$
phase space corresponding to the nucleation of two domain walls.
The unspecified initial noise field $p_0(x)$ acts as an initiator.
The noise profile $p_0(x)$ is localized at the position of the
nucleation. Since the orbit lies on an energy surface we have
$E_0=(1/2)\int dx p_0(x)^2$ and we infer the energy density ${\cal
H}=(1/2)p_0(x)^2$. From the equation of motion $du(x)/dt\approx
p_0(x)$ and inserting in the action in Eq. (\ref{act10}) we obtain
$S_{\text{nucl}}\approx(1/2)\delta t\int dx[p_0(x)+
\Gamma(-\partial^2p_0(x)/\partial x^2+ 4k_0^2p_0(x))\delta
t)p_0(x)-(1/2)p_0(x)^2]$. Rearranging and performing a partial
integration, assuming $p_0(x)$ localized, we find the following
expression for the nucleation action per domain wall:
\begin{eqnarray}
S_{\text{nucl}}\approx\frac{1}{2} \delta tE_0+\frac{1}{2}(\delta
t)^2\left(8k_0^2\Gamma E_0+\Gamma\int dx\left(\frac{\partial
p_0(x)}{\partial x}\right)^2\right).
\end{eqnarray}
This is a short time estimate and holds for a large system, i.e.,
for $k_0L\gg 1$. The energy $E_0=2(1/2)mv^2$ is found from the
2-domain wall sector. Inserting $\delta t=4T/k_0L$, $v=L/2T$,
$m=4k_0/3$, estimating $(\partial p_0(x)/\partial
x)^2=fk_0^2p_0(x)^2$, where $f$ ia a ``fudge'' factor of order
one, and lumping the term linear in $\delta $ with the domain wall
pair propagation, we obtain $S_{\text{nucl}}=\delta
t^2(4+f)k_0^2\Gamma E_0$. Further reduction yields the result
\begin{eqnarray}
S_{\text{nucl}}=(4+f)\Gamma m \label{estim}
\end{eqnarray}
This expression for the nucleation action has the same form as the
one derived from the simple Kramers theory. In Fig.~\ref{fig15} we
have in a phase space plot sketched the orbit and indicated the
nucleation, propagation, and annihilation regions.
\begin{figure}
\includegraphics[width=0.5\hsize]{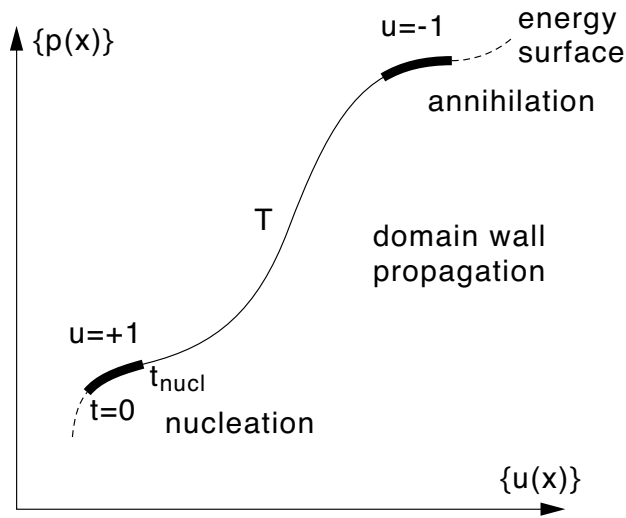}
\caption{We sketch the orbit in phase space corresponding to the
nucleation of a domain wall pair during time $t_{\text{nucl}}$,
the subsequent propagation across the system during time
$T-2t_{\text{nucl}}$, and the final annihilation of the individual
domain wall at the boundaries during time $t_{\text{nucl}}$.}
\label{fig15}
\end{figure}
%
\section{Interpretation of numerical result}
Here we make contact with the numerical analysis of the
Ginzburg-Landau equation by E, Ren, and Vanden-Eijnden \cite{E02}.
These authors analyze the noise induced switching by means of an
optimization techniques applied to the Freidlin-Wentzel action
\begin{eqnarray}
S_{\text{FW}}=\frac{1}{2}\int_0^L dx\int_0^T
dt\left(\frac{\partial u}{\partial t}-\Gamma\frac{\partial^2
u}{\partial x^2}+2\Gamma k_0^2u(1-u^2)\right)^2. \label{fw}
\end{eqnarray}
for a system of size $L=1$ over a time span $T$ and find that the
global minimum of $S_{\text{FW}}$  corresponds to the nucleation
and propagation of domain walls. First we observe that the
minimizing configurations, the minimizers, of (\ref{fw}) are
identical to the orbits found within the canonical phase space
approach. This correspondence was discussed in Sec. III and is
seen by noting that insertion of the equation of motion for $p$
and the Hamiltonian (\ref{ham2}) in the expression for the action
(\ref{act2}) yields the Freidlin-Wentzel form in (\ref{fw}).

Switching a system of size $L$ in time $T$ by means of a single
domain wall, corresponding to the pathway via the lowest local
minimum of the free energy at $F_{\text{dw}}=m$, the propagation
velocity $v=p_0/m=L/T$ and we obtain the action
$S_1(T)=S_{\text{nucl}}+mL^2/2T$ in Eq. (\ref{as1}) and associated
transition probability
\begin{eqnarray}
P\propto\exp\left(-\frac{S_{\text{nucl}}}{\Delta}\right)
\exp\left(-\frac{mL^2}{2T\Delta}\right).
\end{eqnarray}
In the thermodynamic limit $L\rightarrow\infty$, $P\rightarrow 0$
as a result of the broken symmetry in the double well potential.
At long times the action falls off as $1/T$. At intermediate times
$t$ and positions $x$ we have $P\propto\exp(-mx^2/2\Delta t)$ and
the domain wall in the stochastic interpretation perform a random
walk with mean square displacement $2\Delta t/m$, corresponding to
diffusive behavior. In Fig.~\ref{fig12} a we have shown a domain
wall nucleating at the left boundary and propagating with constant
velocity $v=1/T$ to the right boundary, where it annihilates. We
have used the same parameter values as in \cite{E02}, i.e.,
$\delta = \Gamma = .03$, $2\Gamma k_0^2= \delta^{-1}$, $T=7$, and
a system size $L=1$. In Fig.~\ref{fig12} b we have plotted the
trajectory of the domain wall in space and time.

The switching can also take place by nucleating two domain walls
at the boundaries. These then move at half the velocity $v/2$ and
subsequently annihilate at the center. This process corresponds to
the pathway via the local saddle point of the free energy at
$F_{\text{dw}}=2m$, and the action is given by
$S_2(T)=2S_{\text{nucl}}0+2S_1(4T)$ in Eq. (\ref{as2}). The
associated transition probability is
\begin{eqnarray}
P\propto\exp\left(-\frac{2S_{\text{nucl}}}{\Delta}\right)
\exp\left(-\frac{mL^2}{4T\Delta}\right).
\end{eqnarray}
Snapshots of this process are shown in Fig.~\ref{fig13} a and the
corresponding space-time plot in Fig.~\ref{fig13} b.

Combining the contributions from nucleation and the subsequent
domain wall propagation we obtain the heuristic expression in Eq.
(\ref{asn}), where $S_{\text{nuc}}$ is the action for nucleating a
single domain wall and $n$ is the number of walls. From the
improved estimate of $S_{\text{nucl}}$ in Eq. (\ref{estim}) with
fudge factor $f\approx 1$ we have
\begin{eqnarray}
S_{\text{nuc}}\sim 5 \Gamma k_0. \label{snuc}
\end{eqnarray}
In Fig.~\ref{fig16} we have plotted $S$ versus $T$ for $n=1-6$
domain walls using the parameter values in \cite{E02}. Choosing
$S_{\text{nuc}}$ according to (\ref{snuc}) we find excellent
agreement with the numerical results.
\begin{figure}
\includegraphics[width=0.5\hsize]
{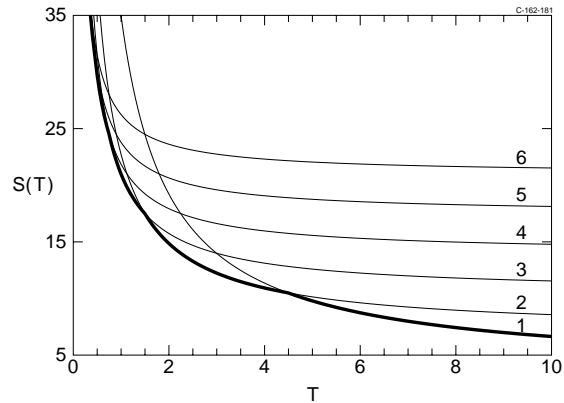}
 \caption{The action $S(T)$ given by
(\ref{asn}) is plotted as a function of $T$ for transition
pathways involving up to $n=6$ domain walls. The lowest action and
thus the most probable transition is associated with an increasing
number of domain walls at shorter times, indicated by the heavy
limiting curve. The curves correspond to choosing
$S_{\text{nuc}}=5\Gamma k_0$.} \label{fig16}
\end{figure}

As also discussed in \cite{E02} we note that the switching
scenario depends on $T$. At shorter switching times it becomes
more favorable to nucleate more domain walls. In the present
formulation this feature is associated with the finite nucleation
or annihilation action $S_{\text{nuc}}$. This is evidently a
finite size effect in the sense that the action at a fixed $T$
diverges in the thermodynamic limit $L\rightarrow\infty$,
corresponding to the broken symmetry.
\section{Summary and Conclusion}
%
In the present paper we have implemented the dynamical phase space
approach developed previously for the noisy Burgers equation to
the case of the one dimensional noise-driven Ginzburg-Landau
equation. Based on a linear analysis of the static domain wall
solutions in the noiseless case we find that the kinetic
transitions take place by means of propagating multi-domain wall
configurations with superimposed local deformation and extended
diffusive modes. The approach also allows for a determination of
the Arrhenius factor associated with the transitions. The
motivation for the present work is a recent numerical optimization
study by E et. al. \cite{E02} of the noisy one dimensional
Ginzburg-Landau equation based on the Freidlin-Wentzel theory of
large deviations. We find excellent agreement both qualitatively
and quantitatively with the numerical finding of E. et al.
\cite{E02}. The dynamical approach offers in the nonperturbative
weak noise or low temperature limit an alternative way of
determining dynamical pathways and the Arrhenius part of the
associated transition rates.

\end{document}